\documentclass{article}

\usepackage{arxiv}
\usepackage{multirow}
\usepackage{threeparttable}
\usepackage[utf8]{inputenc} % allow utf-8 input
\usepackage[T1]{fontenc}    % use 8-bit T1 fonts
\usepackage{hyperref}       % hyperlinks
\usepackage{url}            % simple URL typesetting
\usepackage{booktabs}       % professional-quality tables
\usepackage{amsfonts}       % blackboard math symbols
\usepackage{nicefrac}       % compact symbols for 1/2, etc.
\usepackage{microtype}      % microtypography
\usepackage{lipsum}		% Can be removed after putting your text content
\usepackage{graphicx}
\usepackage{natbib}
\usepackage{doi}
\usepackage{bm}
\usepackage{amsmath} 
\usepackage{xcolor}
\usepackage{graphicx}
\usepackage{subcaption}
\usepackage{comment}
\usepackage{float}
\usepackage{array}
\usepackage{caption}
\usepackage{placeins}  % in your preamble

\usepackage{authblk} % For better author formatting
\title{Semiparametric Dynamic Copula Models for Portfolio Optimization}

\author[1]{Savita Pareek }
\author[2]{Sujit K.~Ghosh }
\affil[1]{Department of Mathematics and Statistics, Auburn University, Auburn, USA}
\affil[2]{Department of Statistics, NC State University, Raleigh, USA\thanks{\texttt{Correspondence: sujit.ghosh@ncsu.edu}}}
\renewcommand{\headeright}{Pareek, S., Ghosh S.K.}
\renewcommand{\shorttitle}{Dynamic Copula Models for Portfolio Optimization}

\hypersetup{
pdftitle={Semiparametric Dynamic Copula Models for Portfolio Optimization},
pdfsubject={q-bio.NC, q-bio.QM},
pdfauthor={Sujit K. Ghosh, Savita Pareek},
pdfkeywords={Checkerboard copula, Markowitz Portfolio optimization, Skewed generalized t- distribution},
}

\begin{document}
\maketitle

\begin{abstract}
The mean-variance portfolio model, based on the risk-return trade-off for optimal asset allocation, remains foundational in portfolio optimization. However, its reliance on restrictive assumptions about asset return distributions limits its applicability to real-world data. Parametric copula structures provide a novel way to overcome these limitations by accounting for asymmetry, heavy tails, and time-varying dependencies. Existing methods have been shown to rely on fixed or static dependence structures, thus overlooking the dynamic nature of the financial market. In this study, a semiparametric model is proposed that combines non-parametrically estimated copulas with parametrically estimated marginals to allow all parameters to dynamically evolve over time. A novel framework was developed that integrates time-varying dependence modeling with flexible empirical beta copula structures. Marginal distributions were modeled using the Skewed Generalized T family. This effectively captures asymmetry and heavy tails and makes the model suitable for predictive inferences in real-world scenarios. Furthermore, the model was applied to rolling windows of financial returns from the USA, India and Hong Kong economies to understand the influence of dynamic market conditions. The approach addresses the limitations of models that rely on parametric assumptions. By accounting for asymmetry, heavy tails, and cross-correlated asset prices, the proposed method offers a robust solution for optimizing diverse portfolios in an interconnected financial market. Through adaptive modeling, it allows for better management of risk and return across varying economic conditions, leading to more efficient asset allocation and improved portfolio performance.
	%The proposed semiparametric model which combines nonparametrically estimated copula with parametrically estimated marginals allows all parameters to dynamically evolve over time making it suitable for predictive inference.
%This article introduces a portfolio optimization framework that combines the skewed generalized t-distribution for modeling asset return marginals with the empirical checkerboard copula to capture dynamic dependence structures. 
%Applied to rolling windows of financial returns from the USA, India and Hong Kong economies, this approach addresses the limitations of traditional models that rely on parametric assumptions. By accounting for asymmetry, heavy tails, and cross-correlated asset prices, the proposed method offers a robust solution for optimizing diverse portfolios in an interconnected financial markets.

\end{abstract}

% keywords can be removed

\section{Introduction}
Portfolio optimization aims to construct investment strategies by selecting an optimal mix of assets to achieve objectives such as maximizing returns, minimizing risk, and ensuring stability. Markowitz’s mean-variance portfolio model (\cite{markowitz1952}) remains foundational, enabling optimal asset allocation based on the risk-return trade-off. However, its reliance on restrictive assumptions about asset return distributions limits its applicability to real-world data, which often exhibit asymmetry, heavy tails, and time varying dependencies.

To address these limitations, advanced strategies have been developed to be used alongside parameteric copula structures. These include maximum Sharpe Ratio (MSR), Global Minimum Variance (GMV), Conditional Value-at-Risk (CVaR) and hierarchical risk parity based strategies. For example, \cite{huang2015timevarying} constructed a time-varying copula with varying window lengths to examine the impact of parametric copula families (e.g., Gaussian, Gumbel, Clayton) on portfolio performance during economic cycles, utilizing Markowitz's variance minimization method. \cite{sahamkhadam2023portfolio} extended portfolio optimization through out-of-sample predictive models, employing AR-GARCH-type processes for marginal asset returns and Vine copulas for joint distributions, focusing on GMV, CVaR, and MSR strategies. \cite{valeyre2023optimal} proposed an optimal trend-following portfolio integrating Markowitz, risk parity, agnostic risk parity, and trend-following strategies. Similarly, \cite{avella2024} evaluated the performance of Markowitz optimization, highlighting the superiority of global minimum variance and mean-variance portfolios over randomly weighted approaches.

Despite these advances, majority of the methods rely on fixed or static dependence structures, overlooking the dynamic nature of financial markets. This work introduces a novel framework that combines time-varying dependence modeling with flexible copula structures using the empirical beta copula, as established by \cite{segers2017}. Marginal distributions are modeled using the Skewed Generalized T (SGT) family, which effectively captures asymmetry and heavy tails (\cite{theodossiou1998sgt}, \cite{Theodossiou2016Skewness}). This semiparametric and dynamic methodology evolves parameters over time, providing an adaptive solution to portfolio optimization that addresses the complexities inherent in real-world financial data.
%All of the above models Despite these advancements, existing approaches often rely on fixed or time-independent copulas and neglect dynamic asset interdependencies. This paper proposes a novel framework integrating time-varying dependence structures and arbitrary dependence modeling using the empirical checkerboard copula, a genuine copula as shown in \cite{lu2023nonparametric}. Furthermore, the approach incorporates marginal distributions from the Skewed Generalized T (SGT) family, which accounts for asymmetry and kurtosis \cite{theodossiou1998sgt}. This methodology provides Semiparametric because we combined parametric SGT with nonparametric EC copula and dynamic because we allowed all the parameters to evolve over time a robust solution to portfolio optimization, addressing the limitations of prior models and adapting to the complexities of real-world financial data.

\section{Methodology}

Portfolio theory provides a quantitative framework for building models of volatile assets. It involves three key steps: identifying the joint distribution and dependence structure of returns, selecting a mathematical model to represent the risk-return trade-off, and solving the model to achieve optimal portfolio outcomes. Consider a portfolio consisting of $m$ asset prices, where the return of each asset is observed over $n$ time points. %For each asset $j$, the return vector is denoted as $\mathbf{R}_j = \{R_{1j}, R_{2j}, \ldots, R_{nj}\}$, representing the returns across $N$ time periods and $\mu_{\bm{R}_j}$ is the mean return of jth asset. Steps to construct the portfolio are outlined.
For each asset $j$, the return vector is denoted as $\mathbf{R}_j = (R_{1j}, R_{2j}, \ldots, R_{nj})^T$, representing the returns across $n$ time periods. The mean return of the $j^{\text{th}}$ asset is given by $\mu_{\bm{R}_j}$. The steps to construct the portfolio are outlined below:

%Let $\mu_j$ be the mean return of asset $j$., we estimate the approximate parameter of sgt distribution for eah of the jth asset  

%\subsection*{Steps for Portfolio Modeling \& Optimization}

\begin{enumerate}
    \item \textbf{Definition of Returns:}  
    The returns are defined as:
    \[
   % R_{tj} = 100 \left( \frac{p_t}{p_{t-1}} - 1 \right) \quad \text{or} 
   \quad R_{tj} = \log\left(\frac{p_{tj}}{p_{{(t-1)}j}}\right),
    \]
    where \(p_{tj}\) denotes the adjusted closing price of the \(j^{\text{th}}\) asset on the \(t^{\text{th}}\) day, with \(t = 1, 2, \ldots, n\) and \(j = 1, 2, \ldots, m\).

   % where $p_{tj}$ is the adjusted closing asset price of jth asset at tth day, \(t = 1, 2, \ldots n\) and \(j = 1, 2, \ldots, m\).

    \item \textbf{Marginal Distribution Estimation and Pseudo-Uniform Transformation:}  
    Estimate the marginal distribution of the returns \(R_{tj}\) and transform each marginal to pseudo-uniform variates:
    \[
    U_{tj} = \hat{F}_j(R_{tj}), \quad t = 1, 2, \ldots n; \, j = 1, 2, \ldots, m,
    \]
    where \(\hat{F}_j\) is the estimated marginal cumulative distribution function using SGT model. More details are provided in 
 Section~\ref{sec:sgt}.%(e.g., using SGT or AR-GARCH models). 
    This step generates an \(n \times m\) matrix of pseudo-uniform marginals.

    \item \textbf{Copula Estimation:}  
    Estimate the empirical beta copula denoted by $\hat{C}_n^\beta(u_1, u_2,\ldots,u_m)$ using the pseudo-uniform variates $\{(U_{t1},\ldots,U_{tm}): t=1,2,\ldots,n\}$. Further details are provided in Section~\ref{sec:emp_beta_copula}. %or other smooth genuine copula estimators.

    \item \textbf{Generation of Random Variates and Reverse Transformation:}
    Generate random variates from the estimated genuine copula and transform them back to the marginal returns using the quantile function or by reversing the method in Step 2:
    \[
    \tilde{R}_{ij}=\hat{F}_j^{-1}(\tilde{U}_{ij}), \quad i = 1, 2, \ldots, N; \, j = 1, 2, \ldots, m,\;\mbox{where}\;(\tilde{U}_{i1},\tilde{U}_{i2},\ldots,\tilde{U}_{im})\stackrel{iid}{\sim} \hat{C}(u_1, u_2,\ldots,u_m)
    \]
    where \(N\) is the Monte Carlo sample size chosen to achieve a desired precision. This step gives a \(N \times m\) matrix of \(\tilde{R}\)'s which can be used to estimate covariance matrix $\Sigma_{\tilde{R}}$ based on the multivariate sample $\{(\tilde{R}_{i1},\ldots,\tilde{R}_{im}): i=1,2,\ldots,N\}$

    \item \textbf{Portfolio Optimization:}  
   Optimize the portfolio with Markowitz's model:
    \[
\min_{\mathbf{w}} \mathbf{w}^T \Sigma_{\tilde{R}} \mathbf{w}, \;\;\text{subject to:}
\]
\begin{itemize}
\item   \( \sum_{j=1}^{m} w_j = 1 \)\quad (Full investment, weights sum to 1)
\item  \( w_j \geq 0\), $j=1, 2, \ldots m$ \;\; (Long-only, non-negativity constraint)
\item  \( \sum_{j=1}^{m} w_j \mu_{\bm{R}_j} \geq \frac{1}{m} \sum_{j=1}^{m} \mu_{{\bm{R}}_j}\)\; (equal-weight outperformance (EWO), optimal weighted return is greater or equal the equal-weighted return).

\end{itemize}
In this step, alternative optimization strategies such as MSR, CVaR, or VaR can also be employed. 
    The portfolio weights \(\hat{w}_j\), are estimated by solving quadratic programming optimization problem described above. 
    %using the matrix of \(\tilde{U}\)'s to compute the chosen optimization criterion.

    \item \textbf{Portfolio Net Worth Value and Sharpe Ratio Computation:}  
The portfolio's net worth and Sharpe ratio are computed using the next day's expected returns, covariance matrix, and the current optimal portfolio weights. These weights, $\hat{\bm{w}}$, are obtained by repeating Steps 1 to 5 on the current rolling return matrix, $\bm{R}_{(t)}$, constructed based on a specified rebalancing frequency.

\begin{align*}
\text{Net Worth} &= \hat{\bm{w}}^\prime E(\bm{R}_{(t+f)}), \\
\text{Sharpe Ratio} &= \frac{\hat{\bm{w}}^\prime E(\bm{R}_{(t+f)})}{\sqrt{\hat{\bm{w}}^\prime \text{Cov}(\bm{R}_{(t+f)}) \hat{\bm{w}}}},
\end{align*}
where
\[
\bm{R}_{(t+f)} = 
\begin{bmatrix}
\bm{R}_{t+f,1} & \bm{R}_{t+f,2} & \cdots & \bm{R}_{t+f,m}
\end{bmatrix}^\prime
\]
denotes the vector of asset returns for $m$ assets on the $(t+f)^{\text{th}}$ day, corresponding to the next rebalancing point. Assuming a rolling window of length $L$ and a rebalancing frequency denoted by $f$, the rolling return matrix $\bm{R}_{(t)}$ includes daily returns from the $t^{\text{th}}$ day to the $(t+L-1)^{\text{th}}$ day. The return vector $\bm{R}_{(t+f)}$, used for evaluating portfolio performance at the next rebalancing step, spans the interval from the $(t+f)^{\text{th}}$ day to the $(t+L+f-1)^{\text{th}}$ day.
We consider $f=1$ (daily) and $f=5$ (weekly) rebalancing, though the framework generalizes to any frequency (e.g., fortnightly $f=10$, monthly $f=21$) as required by the application.

    \item \textbf{Comparison of Portfolio Net Worth and Sharpe Ratio:}  
    %Compare the portfolio net worth value obtained with other weights, such as Equal-weight portfolio or using other optmisation strategy as GMV, CVaR etc..
   Compare the portfolio's next day net worth, Sharpe ratio with results obtained using equal-weighted portfolios or alternative optimization strategies.% such as GMV and CVaR.

    \item \textbf{Performance Plots:}  
    Generate plots for: (a) Empirical rolling average returns in a portfolio, (b) Empirical rolling standard deviations in a portfolio (c) Rolling optimal weights using a specific optimization criterion, and (d) Compare next day portfolio performance with other methods.
  %  \begin{enumerate}
      %  \item Empirical rolling average returns in a portfolio
       % \item Empirical rolling Sharpe ratio %Value-at-Risk (VaR), or Conditional Value-at-Risk (CVaR)  of returns in a portfolio
       % \item Rolling optimal weights using a specific criterion
       % \item Compare portfolio performance with other methods
   % \end{enumerate}
\end{enumerate}

Next, we describe the details of each of the key steps above in our models.

\subsection{Marginal Density Modeling using Skewed Generalized t Distribution}
\label{sec:sgt}
%\citet{theodossiou1998sgt} as an extension of the \textit{Generalized T distribution}. 
The Skewed Generalized T (SGT) distribution introduced by \cite{theodossiou1998sgt} is a five-parameter distribution and a flexible methodology to model financial data with asymmetric distributions and heavy tails. %that models data with heavy tails and skewness, making it useful for financial returns and risk analysis. %is
Due to its ability to account for skewness and excess kurtosis, the SGT is particularly effective at capturing the true distribution of asset returns, exchange rates, and commodity prices. \cite{theodossiou1998sgt} provides empirical evidence that this model represents financial risks with a better fit than symmetrical models.  In addition, \cite{Theodossiou2016Skewness} demonstrate that skewness plays a key role in explaining inconsistencies in the risk-return relationship, demonstrating that models that assume symmetry underestimate risk exposure.

Many well-known heavy-tailed and skewed distributions are special cases of the SGT distribution, including the Skewed Generalized Error, Generalized T, Skewed T, Skewed Laplace, Generalized Error, Skewed Normal, Student's T, Skewed Cauchy, Laplace, Uniform, Normal, and Cauchy distributions. The probability density function (PDF) of the SGT distribution is defined as follows.

A random variable \(X\) is said to follow the SGT distribution, denoted \(X \sim SGT(\mu, \sigma, \lambda, p, q)\), if its PDF is given by:
\[
f_{SGT}(x; \mu, \sigma, \lambda, p, q) = 
\frac{p}{2 \nu \sigma q^{1/p} B\left(\frac{1}{p}, q\right)} 
\left( \frac{|x - \mu + m|^p}{q (\nu \sigma)^p (\lambda \operatorname{sign}(x - \mu + m) + 1)^p} + 1 \right)^{-\frac{1}{p} - q},
\]
where \(x \in \mathbb{R}\), and the parameters satisfy \(\mu \in \mathbb{R}, \; \sigma > 0\), \(-1 < \lambda < 1\), and \(p, q > 0\). The parameter \( \mu \) defines the central location of the distribution, while \( \sigma \) determines its scale. The skewness of the distribution is controlled by \( \lambda \), whereas \( p \) and \( q \) jointly regulate its kurtosis.
 
 When \(\lambda = 0\), the distribution is symmetric. For \(-1 < \lambda < 0\), it is negatively skewed, while \(0 < \lambda < 1\) results in positive skewness.
Smaller values of \(p\) and \(q\) produce a leptokurtic distribution with heavy tails, whereas larger values yield a platykurtic distribution with lighter tails. The \(h\)-th moment of the SGT distribution exists only if \(pq > h\), indicating that the finiteness of higher-order moments depends on the choice of these parameters. 

The terms \(m\) and \(v\) are defined as:
\[
m = \frac{2 \nu \sigma \lambda q^{1/p} B\left(\frac{2}{p}, q - \frac{1}{p}\right)}{B\left(\frac{1}{p}, q\right)},
\]
%if \texttt{mean.cent = TRUE}, and \(m = 0\) otherwise. Note that \(pq \leq 1\) with \texttt{mean.cent = TRUE} is an error, and \texttt{NaNs} will be produced.

\[
v = q^{-\frac{1}{p}} \left[ 
\left(3\lambda^2 + 1\right) 
\frac{B\left(\frac{3}{p}, q - \frac{2}{p}\right)}{B\left(\frac{1}{p}, q\right)} 
- 4\lambda^2 \left( 
\frac{B\left(\frac{2}{p}, q - \frac{1}{p}\right)}{B\left(\frac{1}{p}, q\right)}
\right)^2 
\right]^{-\frac{1}{2}},
\]
%if \texttt{var.adj = TRUE}. If \texttt{var.adj = FALSE}, then \(v = 1\). Note that \(pq \leq 2\) with \texttt{var.adj = TRUE} is also an error, and \texttt{NaNs} will be produced.

where the beta function is defined as
\(
B(a, b) = \frac{\Gamma(a)\Gamma(b)}{\Gamma(a + b)},
\)
with \( \Gamma(\cdot) \) denoting the gamma function. This study utilizes the SGT distribution to model financial return data with asymmetry and fat tails. Its flexibility in capturing skewness and excess kurtosis makes it ideal for such data. By applying a rolling window approach, we fit the SGT distribution dynamically, allowing for time-varying analysis and improved modeling of financial market dynamics.

\subsection{Nonparametric Copula Model Using Smoothed Beta Copula}
\label{sec:emp_beta_copula}
Copula models are increasingly used in financial time series analysis to capture the evolving dependence structure in asset returns and stock price movements. Their ability to model complex dependencies makes them a fundamental tool for analyzing multivariate data, as they link univariate distribution functions into a multivariate framework. By Sklar’s theorem (\citet{sklar1959}), any joint distribution can be uniquely expressed in terms of its marginal distributions and a copula, which encapsulates the dependence structure among multiple random variables. This decomposition enables flexible estimation, allowing marginals and the copula to be modeled separately using distinct techniques (\citet{jaworski2010}, \citet{patton2012}, \citet{joe2014}).
For a portfolio return vector of $d$ assets \(({X}_1, \ldots, {X}_d)\) with a joint cumulative distribution function (CDF) \(F\) and continuous marginal CDFs \(F_j, \, j = 1, \ldots, d\), Sklar’s theorem asserts that:
\[
F(x_1, \ldots, x_m) = C(F_1(x_1), \ldots, F_d(x_d)),
\]
where \(C(\cdot)\) is the copula function. A copula is the joint CDF of a transformed random vector \((U_1 = F_1(X_1), \ldots, U_d = F_d(X_d))\), where the marginal distributions are uniform on \([0,1]\). While Sklar’s theorem applies to discrete variables, this study focuses on continuous multivariate random vectors (e.g., stock returns), assuming all marginal CDFs \(F_j\) are absolutely continuous.

Copulas can be estimated through parametric or nonparametric methods. Parametric models, such as Gaussian, t, or Archimedean copulas (\citet{nelsen2007}, \citet{mcneil2009}), are computationally efficient and interpretable but may fail to capture complex dependencies like asymmetry or tail dependence, especially under model misspecification. Vine copulas address some of these challenges by constructing dependence structures using bivariate copulas at each node of a vine tree. However, their estimation often involves high-dimensional integrals, posing computational challenges. Nonparametric approaches, including empirical, Bernstein, and beta copulas, provide greater flexibility in capturing arbitrary dependencies (\citet{sancetta2004}, \citet{genest2017}, \citet{segers2017}). However, many nonparametric estimators are valid copulas only asymptotically, limiting their applicability to finite samples. Furthermore, dependence measures like Spearman’s rho and Kendall’s tau, based on these estimators, can fall outside their natural range, making them impractical in some scenarios. To overcome these limitations, \citet{segers2017} introduced the empirical beta copula (EBC), a valid, smooth copula constructed as a special case of the empirical Bernstein copula with polynomial degrees equal to the sample size. It improves bias and mean squared error over classical estimators, but may have higher variance than Bernstein copulas with lower degrees. \citet{lu2023nonparametric} extended this idea via the empirical checkerboard Bernstein copula (ECBC), allowing for varying polynomial degrees. In this work, we adopt the EBC due to its greater computational efficiency compared to ECBC. The EBC is defined as:

\begin{equation*}
C_n^{\beta}(\boldsymbol{u}) = \frac{1}{n} \sum_{i=1}^n \prod_{j=1}^d F_{n, R_{i,j}^{(n)}}(u_j), \quad \boldsymbol{u} = (u_1, \ldots, u_d) \in [0,1]^d, 
\end{equation*}
 where \(R_{i,j}^{(n)}\) denotes the rank of \(X_{ij}\) among \((X_{1j}, \dots, X_{nj})\). For $u \in [0,1]$ and $r \in \{1, \ldots, n\}$,
\begin{equation*}
F_{n, r}(u) = \mathbb{P}(U_{r:n} \leq u) = \sum_{s=r}^n \binom{n}{s} u^s (1 - u)^{n - s}
\end{equation*}
is the cumulative distribution function of beta distribution $\mathcal{B}(r, n + 1 - r)$. Here, $U_{1:n} < \cdots < U_{n:n}$ denote the order statistics based on $n$ independent random variables $U_1, \ldots, U_n$ uniformly distributed on $[0,1]$.
\begin{comment}
\[
C_n^\#(u) =
\frac{1}{n} \sum_{i=1}^n \prod_{j=1}^d \min\left(\max\left(n u_j - R_{i,j}^{(n)} + 1, 0\right), 1\right),
\]
where \(R_{i,j}^{(n)}\) denotes the rank of \(X_{ij}\) among \((X_{1j}, \dots, X_{nj})\). A smooth copula estimator, \(C_{m,n}^\#(u)\), is constructed from \(C_n^\#(u)\) as:
\[
C_{m,n}^\#(u) = \sum_{k_1=0}^{m_1-1} \cdots \sum_{k_d=0}^{m_d-1} \tilde{\theta}_{k_1, \dots, k_d} \prod_{j=1}^d \binom{m_j}{k_j} u_j^{k_j} (1 - u_j)^{m_j - k_j},
\]
where \(m_1, m_2, \ldots, m_d\) are the polynomial degrees for each dimension, and \(\tilde{\theta}_{k_1, \dots, k_d}\) are coefficients defined as:
\[
\tilde{\theta}_{k_1, \dots, k_d} = C_n^\#\left(\frac{k_1}{m_1}, \dots, \frac{k_d}{m_d}\right)=
\tilde{\theta}_{k_1, \dots, k_d} =
\frac{1}{n} \sum_{i=1}^n \prod_{j=1}^d \min\left(\max\left(n \frac{k_j}{m_j} - R_{i,j}^{(n)} + 1, 0\right), 1\right),
\]
where \(k_j \in \{0, 1, \ldots, m_j - 1\}\) for \(j = 1, 2, \ldots, d\).
\end{comment}

We employ the EBC to estimate time-varying copulas over rolling windows of financial returns. This approach is particularly useful for capturing dynamic dependence structures, as dependencies between assets often evolve during periods of market turbulence, such as financial crises. %By allowing the dependence structure to vary over time, the ECBC provides a flexible tool for modeling dynamic relationships in financial data. 
We implement the EBC using the \texttt{empCopula} function from the \texttt{copula} package in R, with smoothing set to \texttt{beta} 
 (\citet{copula}). This method is further integrated into a semiparametric framework, where the marginals are modeled using the Skewed Generalized T distribution, enabling us to capture asymmetry, heavy tails, and time-varying dependencies in financial markets. %\textcolor{red}{?check as the package says they incorporate the empirical checkerboard copula based on  Segers, Sibuya and Tsukahara (2017): GHOSH COMMENT: From the copula package manual, it appears that you are not using the smoothed copula and you are using eq. (4.1) of the SST(2017) paper which is empirical checkerboard copula. The equation that we have in our paper is akin to eq. (2.1) in their paper which is the "beta" option in the copula package.}

\subsection{Portfolio Optimization using Quadratic Programming}
Optimizing portfolios and diversifying investments have been crucial in financial decision making. Markowitz's mean variance optimization model (MVO) (\citet{markowitz1952}) introduced a quantitative framework to balance risk and return, formulating portfolio selection as an optimization problem. It emphasizes diversification, where risk depends on asset correlations rather than individual asset risk. Using the model, investors select portfolios with the lowest variance and reject inefficient portfolios with higher risk. This approach revolutionized classical financial analysis by shifting the focus from single-asset valuation to portfolio-level risk management. The MVO framework is typically formulated as a quadratic programming problem, where investors maximize expected returns for a given risk level or minimize portfolio variance for a required return. For a detailed exposition and practical considerations in portfolio optimization, refer to \citet{Bacon2008}, \citet{Kolm2014}, and \citet{Palomar2025}. Let $d$ be the number of assets in an investment universe with uncertain future returns denoted by
\( \mathbf{r} = (r_1, r_2, \dots, r_d)^T \). A portfolio is represented by the weight vector 
\( \boldsymbol{\omega} = (\omega_1, \omega_2, \dots, \omega_d)^T \), where \( \omega_i \) is the proportion 
of total funds allocated to asset \( i \). The portfolio return is given by:
\[ 
r_p(\boldsymbol{\omega}) = \boldsymbol{\omega}^T \mathbf{r}.
\]
The expected portfolio returns and  standard deviation (risk) of the portfolio return are represented as:
\[
{\mu(\boldsymbol{\omega})}=\boldsymbol{\mu}^T \boldsymbol{\omega},\quad
\sigma(\boldsymbol{\omega}) = \sqrt{\boldsymbol{\omega}^T \boldsymbol{\Sigma} \boldsymbol{\omega}}.
\]
where:
\begin{itemize}
    \item \(\boldsymbol{\mu} = (\mu_1, \mu_2, \dots, \mu_d)^T\),\;
    \( \mu_i = \mathbb{E}(r_i) \) is the expected return of asset \( i \).
    \item \( \boldsymbol{\Sigma} = \operatorname{diag}(\boldsymbol{\sigma}) \boldsymbol{P} \operatorname{diag}(\boldsymbol{\sigma}) \) is the covariance matrix, which is positive semi-definite, ensuring \( \boldsymbol{\omega}^T \boldsymbol{\Sigma} \boldsymbol{\omega} \geq 0 \). 
%\item \(\boldsymbol{\Sigma} = \operatorname{diag}(\boldsymbol{\sigma}) \boldsymbol{P} \operatorname{diag}(\boldsymbol{\sigma}),\)
    \item \( \boldsymbol{\sigma} = (\sigma_1, \sigma_2, \dots, \sigma_d)^T \) is the vector of standard deviations.
    \item \( \boldsymbol{P} = [\rho_{ij}] \) is the correlation matrix, where \( \rho_{ij} \) is the correlation between asset \( i \) and \( j \).  
\end{itemize}

Let \( \Omega \subset \mathbb{R}^d \) denote the set of permissible portfolios, where \( \boldsymbol{\omega} \in \Omega \) satisfies the portfolio constraints. Using this framework, the MVO problem is formulated as:
\[
\max_{\boldsymbol{\omega} \in \Omega}  \boldsymbol{\mu}^T \boldsymbol{\omega} - \lambda \boldsymbol{\omega}^T \boldsymbol{\Sigma} \boldsymbol{\omega},\quad
\text{subject to:}\;\; \mathbf{1}^T \boldsymbol{\omega} = 1, \;\; \boldsymbol{\omega} \geq 0,
\]
where \( \lambda \) is the investor-specific risk-aversion parameter governing the trade-off between expected portfolio return and portfolio risk. When \( \lambda = 0 \), the optimization problem focuses solely on maximizing expected return, resulting in the global maximum return portfolio. Conversely, as \( \lambda \to \infty \), the problem minimizes risk entirely, producing the GMV portfolio. 
Alternative formulations of portfolio optimization include the MSR portfolio, as well as approaches based on alternative risk measures such as downside risk, semivariance, VaR, CVaR, and drawdown (\citet{Gunjan2023}).

In this study, we implement the GMV portfolio strategy using a covariance matrix estimated from copula-based simulated returns. To ensure a reasonable level of predictive performance, the optimization is subject to long-only, full investment, and an additional EWO return constraint requiring the portfolio's expected return to exceed the average return of the individual assets. The optimization problem is stated as:
\[
\min_{\boldsymbol{\omega} \in \Omega} \quad \boldsymbol{\omega}^\top \boldsymbol{\Sigma} \boldsymbol{\omega}, \quad
\text{subject to:} \quad
\mathbf{1}^\top \boldsymbol{\omega} = 1,\quad 
\boldsymbol{\omega} \geq 0,\quad 
\boldsymbol{\mu}^\top \boldsymbol{\omega} \geq \bar{r},
\]
where \( \bar{r} = \frac{1}{d} \sum_{i=1}^{d} \mu_i \) denotes the average expected return of the \(d\) assets, \( \boldsymbol{\mu} \) is the vector of expected asset returns, and \( \boldsymbol{\Sigma} \) is the covariance matrix estimated from the copula-implied joint return distribution. We solve this optimization problem dynamically over rolling windows using the \texttt{solve.QP} function from \texttt{quadprog} package in \texttt{R} (\citet{quadprog2019}). By obtaining optimal weights for each window, we capture changes in market dynamics and improve portfolio performance over time. 

The proposed portfolio optimization framework introduces a flexible  structure by combining three components: (1) nonparametric copula-based modeling of dependence, (2) the inclusion of EWO return constraint to ensure practical applicability, and (3) a rolling window of approximately one year of daily returns to dynamically adjust portfolio weights. The first component addresses the limitations of commonly used parametric copulas, such as Gaussian and T copulas, which can capture certain nonlinear and tail dependencies but are limited by their fixed functional forms. These limitations become more pronounced in high-dimensional settings or when the true dependence structure is complex or asymmetric. In contrast, nonparametric copulas provide greater flexibility by allowing the joint dependence structure to be estimated directly from the data without imposing a specific parametric form (\citet{nelsen2007}, \citet{patton2012}). This is particularly advantageous when asset returns exhibit non-Gaussian features, as is commonly observed in financial markets. The second component, an EWO return constraint, ensures that the optimized portfolio achieves an expected return at least as high as that of a naive equal-weighted portfolio. This not only aligns the model with realistic investor expectations, but also safeguards against low-return allocations that may arise in purely risk-driven optimization settings. The third component incorporates a rolling window framework (typically 250 trading days, adjusted for the specific market’s calendar) with daily or weekly rebalancing to account for changing market dynamics, producing real-time optimal allocations that respond to structural shifts in the joint return distribution.

%ensures that portfolios are not only risk-efficient but also achieve returns exceeding market benchmarks, thus enhancing practical relevance.
A key theoretical motivation for this framework lies in the inadequacy of assuming multivariate normality. When asset marginals are normally distributed, the joint distribution is also multivariate normal, and the covariance matrix fully characterizes the dependence structure. However, empirical evidence consistently shows that financial returns exhibit skewness, leptokurtosis, and heavy tails, often due to volatility clustering, price jumps, and higher-moment dependencies (\citet{Bali2007SGTVaR}). When marginals follow skewed distributions such as SGT or skew-normal, there is no natural or unique multivariate extension that preserves the given marginals while maintaining analytical tractability (\citet{theodossiou1998sgt}, \citet{Azzalini2003}). Moreover, the covariance structure cannot be inferred directly from these marginals, making mean-variance optimization unsuitable. In contrast, the copula-based approach allows one to model such joint behaviors by specifying each marginal independently and linking them through a copula, enabling simulation of artificial datasets that reflect the empirical joint behavior. Although this approach has a small computational overhead, it offers significantly improved realism and predictive accuracy. 
While our approach is anchored in the classical mean-variance paradigm, it can be extended to accommodate higher-order moments or incorporate quantile-based risk measures, such as VaR and CVaR, to better reflect tail-risk behavior in non-Gaussian return distributions.% \citep{Rockafellar2000CVaR, Harvey2010higher}.

\subsection{Visualization of Portfolio Performance Measures}
This section analyzes the time-varying performance of two U.S. assets, Netflix (NFLX), a volatile stock, and Costco (COST), a stable stock, using daily log returns from April 2018 to March 2025 with weekly rebalancing. We examine rolling mean returns, standard deviations, Sharpe ratios, correlations, and return densities to capture their dynamic risk-return profiles and regime shifts over time.

%Figure \ref{fig:roll_mean1}(a) illustrates the dynamic nature of the rolling mean returns, which fluctuate between approximately -0.004 and 0.004. NFLX exhibits greater variability than COST (which is in the range -.001 to 0.002), indicating higher sensitivity to market fluctuations. Figure \ref{fig:roll_mean1}(b) presents the rolling standard deviation, where NFLX consistently demonstrates greater volatility, with values ranging from 0.02 to 0.05. The time variation in volatility highlights alternating periods of increased and decreased risk exposure. Figure \ref{fig:roll_mean1}(c) shows that the rolling Sharpe ratio remains consistently low, generally below 0.2, implying minimal risk adjusted returns for both stocks. Figure \ref{fig:roll_mean1}(d) 

\begin{figure}[htbp]
    \centering
    \includegraphics[width=0.8\textwidth]{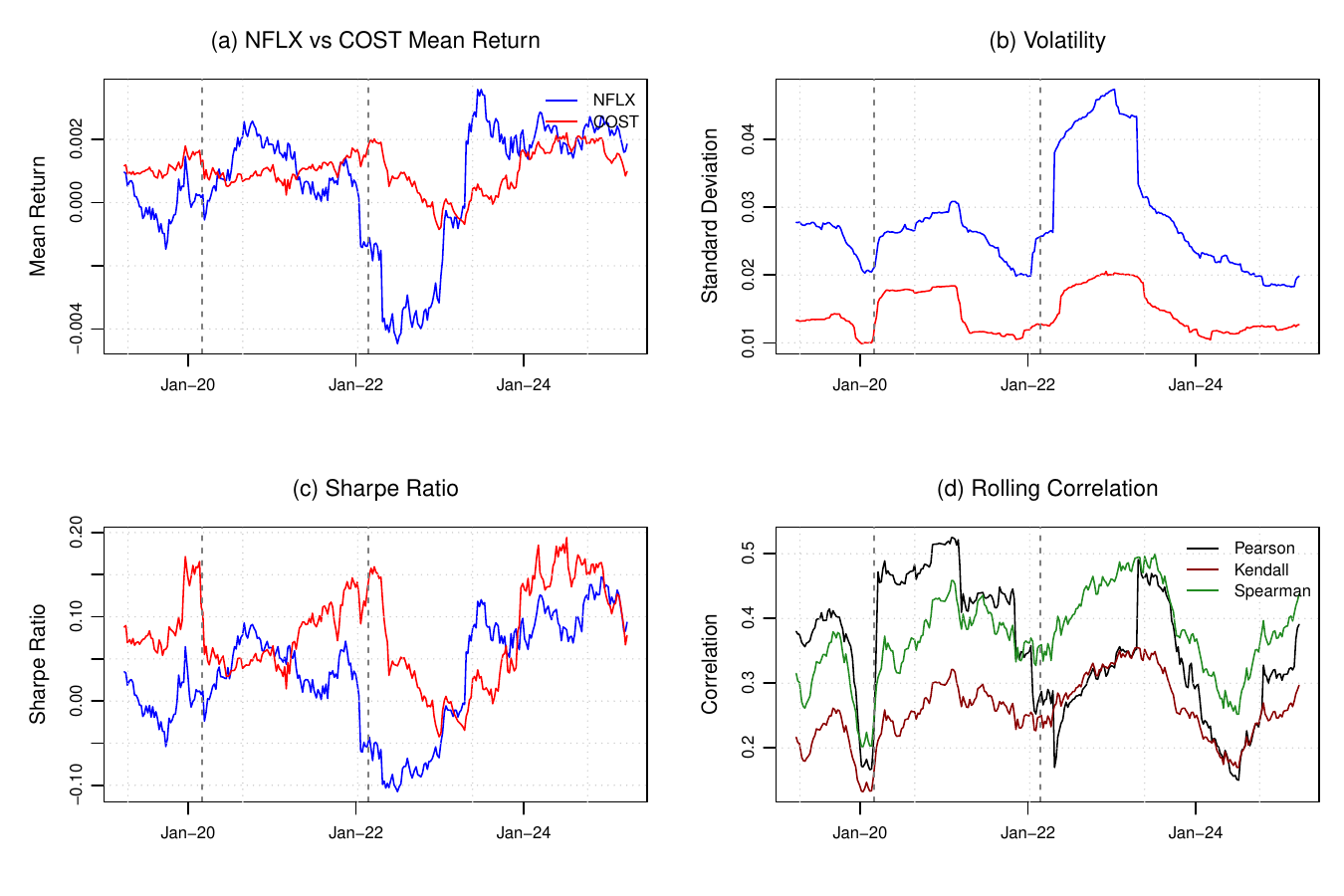} % Replace with your file name
   \caption{Rolling performance metrics for NFLX and COST based on log Returns (250-Day window, weekly rebalancing): 
Plot (a) represents the rolling expected portfolio return, (b) shows the rolling standard deviation, and (c) illustrates the rolling Sharpe ratio, defined as the ratio of mean return to standard deviation. 
Plot (d) depicts the rolling correlation using Spearman's rank, Kendall's tau, and Pearson correlation, capturing the time-varying relationships between NFLX and COST assets.}

    \label{fig:roll_mean1}
\end{figure}

Figure~\ref{fig:roll_mean1} provides a comprehensive view of the evolving return dynamics between NFLX and COST. Panel~(a) displays the rolling mean returns, highlighting pronounced fluctuations for NFLX (ranging from approximately -0.004 to 0.004), while COST exhibits a narrower band (approximately -0.001 to 0.002), consistent with its lower market sensitivity. Panel~(b) shows the rolling standard deviations, where NFLX consistently exhibits higher volatility (0.02 to 0.05), depicting its riskier profile. Panel~(c) presents the rolling Sharpe ratios, which remain generally below 0.2 for both stocks, indicating modest risk-adjusted performance over time. Panel~(d) illustrates the time-dependent relationship between NFLX and COST through rolling correlations. The correlation strength fluctuates between 0.2 and 0.5, reflecting varying co-movement influenced by market conditions. Pearson correlation indicates a strong yet time-varying linear dependence, while Spearman and Kendall correlations exhibit relative stability, suggesting consistent rank ordering despite fluctuations in absolute returns. Kendall’s $\tau$ correlation is lower than Spearman’s $\rho$ due to its stricter pairwise ranking comparison, making it more sensitive to minor fluctuations, whereas Spearman captures broader monotonic relationships. 

%illustrates rolling correlations, with Pearson values varying between 0.2 and 0.5, reflecting dynamic linear co-movement driven by shifting market regimes. In contrast, Spearman and Kendall correlations demonstrate greater temporal stability, suggesting that the relative ranking of returns remains consistent even amid volatility. Notably, Kendall’s $\tau$ remains below Spearman’s $\rho$ due to its stricter treatment of pairwise rank concordance, rendering it more sensitive to small perturbations in the return structure. Together, these results emphasize the asymmetric dependence structure between NFLX and COST, which has implications for diversification and risk management.

\begin{figure}[htbp]
    \centering
    \includegraphics[width=0.8\textwidth]{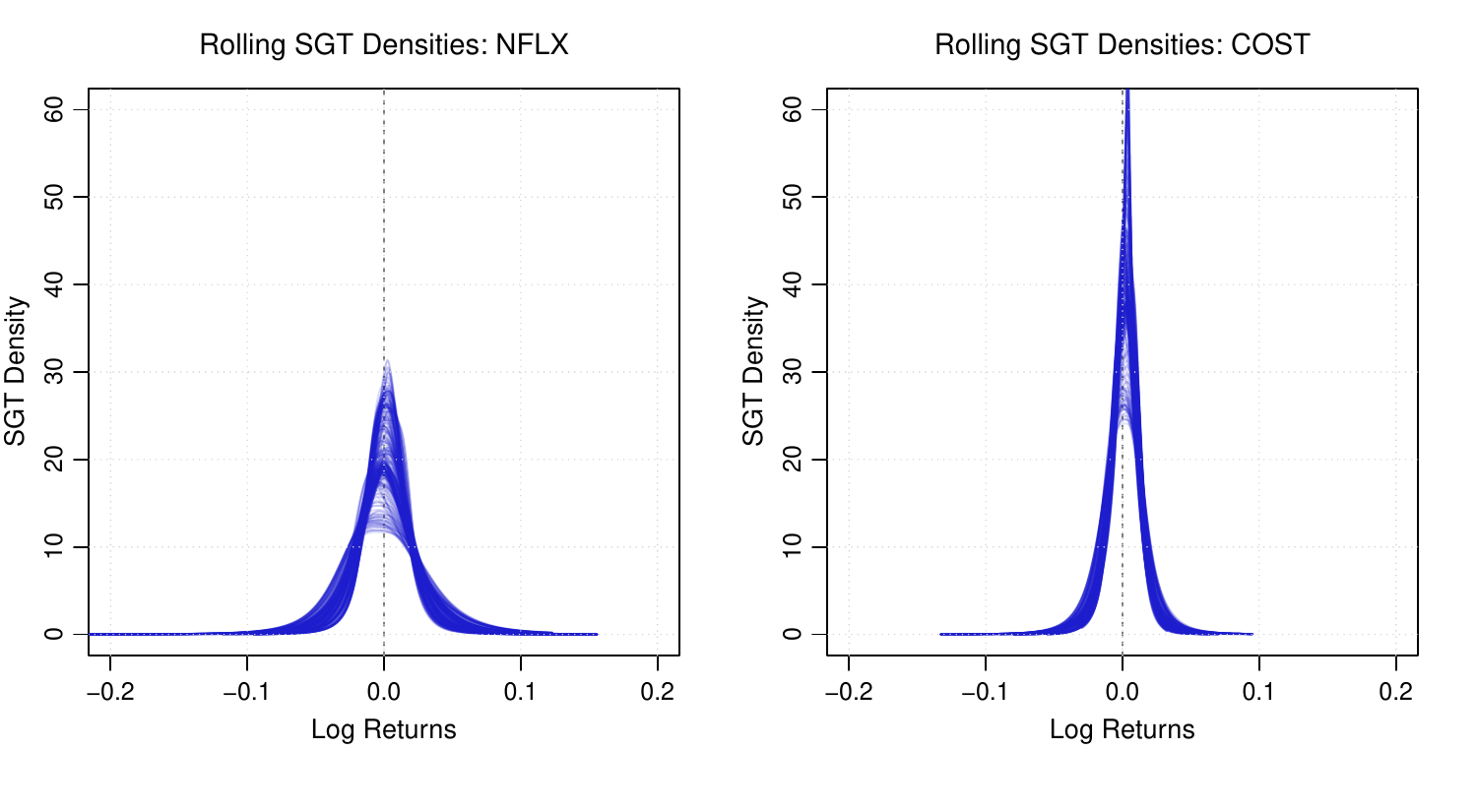} % Replace with your file name
   \caption{Overlaid SGT-fitted density plots for the log-differenced adjusted returns of NFLX and COST over a 250-day rolling window with weekly rebalancing. The x-axis represents log returns, while the y-axis depicts the corresponding estimated SGT densities. The comparison highlights differences in tail behavior and asymmetry between a high-volatility (NFLX) and low-volatility (COST) asset.}

    \label{fig:roll_den1}
\end{figure}

Figure \ref{fig:roll_den1} further reinforces the dynamic nature of stock returns through rolling density plots. For COST, high peaks suggest that returns are concentrated around the mean, indicating lower variability, while lower peaks signal periods of increased volatility. NFLX, on the contrary, exhibits wider and flatter distributions, reflecting greater variance and market sensitivity. These evolving density patterns suggest that the return dynamics is influenced by market conditions and investor sentiment, although the changes are not drastic. The adaptability of the SGT distribution effectively captures high kurtosis and skewness, making it a reliable model for financial returns.

In summary, these findings demonstrate that stock returns, volatility, and correlations evolve dynamically over time, underscoring the interconnected nature of financial markets. Market shocks, investor behavior, and historical return patterns play a crucial role in shaping future movements, highlighting the importance of rolling-window analyses over static measures to better assess financial dependencies.
%\vspace{-2.2cm}

%(iii) rolling density / rolling sharpe ratio / boxplot of rolling optimal weights

\section{Financial Data}
%Additionally, we examine how optimal asset weights are computed over rolling windows of 250 days using data spanning two years. This diverse market selection enables a robust evaluation of portfolio performance across varying economic contexts.
%To assess the proposes stepwise procedure we have selected the developing country India whose returns are more volatile and the developed country USA whose returns are relatively more stable. To check whether the optimal weighted portfolio gives more information capturing market dynamics and more net worth and how the optimal asset weights are computed over rolling windows of length 250 days. The data spans 2 years 
We evaluate the proposed portfolio optimization framework using stock price data from three major markets: the United States (developed), India (developing), and Hong Kong (representing the Far East Asian financial hub). The empirical analysis spans April 1 2018 to March 31st 2025, capturing both stable periods and episodes of extreme market stress, including the COVID-19 crisis. 
Daily equity prices are retrieved from Yahoo Finance via the \texttt{quantmod} package in \texttt{R} \citet{quantmod}. For each asset, we extract adjusted closing prices using the \texttt{Ad()} function and compute log returns as \texttt{diff(log(Ad(get(ticker)))}), ensuring consistency by accounting for corporate actions such as dividends and stock splits. To capture evolving market dynamics and time-varying dependencies, we implement a rolling window approach based on one trading year, with window lengths aligned to local exchange calendars.

Each portfolio comprises 20 large-cap companies selected from the S\&P 500 (United States), NIFTY 50 (India), and the top-traded stocks by turnover listed in HKEX (Hong Kong), chosen for their substantial market capitalization and their significant influence on their respective economies, providing a representative and diversified portfolio for analysis. 
%ensuring sectoral diversity across Information Technology, Financials, Energy, Consumer Staples, and Communication Services. 
A summary of the rolling window setup and average volatility is provided in Table~\ref{tab:rolling-summary}, with asset-level details in Appendix Tables \ref{tab:us-summary-stats}, \ref{tab:india-summary-stats}, and \ref{tab:hk-summary-stats}. While the proposed framework is readily extendable to higher-dimensional portfolios, we focus on 20 assets per market to maintain clarity in visualization and facilitate interpretable inference. The seven-year sample period offers a comprehensive setting to evaluate the procedure’s adaptability across a range of economic environments, including episodes of structural shifts and increased market volatility.

\begin{table}[htbp]
\centering
\caption{Summary of Rolling Window (RW) Setup and Market Characteristics}
\label{tab:rolling-summary}
\begin{tabular}{lcccccccc}
\toprule
{Economy} & {Daily Returns} & {RW Size} & {No. of RW (Daily)} & {No. of RW (Weekly)} & {Avg Volatility (SD, \%)}  \\
\midrule
United States & 1,758 & 250 & 1,508 & 301 & 2.01 \\
India          & 1,728 & 245 & 1,483 & 296 & 1.80  \\
Hong Kong      & 1,719 & 244 & 1,475 & 295 & 1.72\\
\bottomrule
\end{tabular}
\end{table}

\section{Empirical Analysis}
%We evaluated the proposed portfolio optimization framework using data from three markets: India, representing a volatile developing economy, and the USA, a stable developed economy and Hong Kong. The analysis spans January 2022 to January 2024 with a rolling window of 250 days, incorporating ten companies from each market across diverse sectors such as technology, banking, and consumer goods.
We apply the proposed stepwise procedure to the selected markets to evaluate its empirical performance. Using seven years of daily equity data, we compute optimal portfolio weights within rolling windows of 250 trading days for the U.S., 245 for India, and 244 for Hong Kong. The analysis is conducted under both daily and weekly rebalancing schemes, where portfolio weights are updated accordingly, every trading day in the daily scheme, and every five trading days in the weekly scheme. This setup enables us to assess the adaptability of the procedure across varying market conditions and rebalancing frequencies.

Multivariate normality of asset returns was assessed using the \texttt{fPortfolio} package (\cite{fPortfolio}), with results uniformly rejecting the null hypothesis across all economies (\( p\) values \(< 2.2 \times 10^{-16} \)).
The marginal distributions of the asset returns are modeled using the SGT distribution, with parameters estimated via the \texttt{sgt.mle} function. Goodness-of-fit is assessed using the Anderson-Darling (AD) test, with all \( p \)-values exceeding 0.2, indicating adequate fit. 
Dependence structures across assets were modeled using the empirical beta copula (\texttt{empCopula}), which flexibly captures non-Gaussian dependencies and tail behavior. To simulate joint return scenarios from the fitted copula, we employed Monte Carlo sampling with \(10^5\) replications per market. Monte Carlo sample sizes were chosen to ensure target numerical accuracies of \( \varepsilon = 8 \times 10^{-4},\ 7 \times 10^{-4},\) and \(5 \times 10^{-4} \) for the U.S., India, and Hong Kong, respectively. These targets control the simulation error in the empirical copula approximation. The required number of replications was determined using Proposition~1 in \citet{lu2023nonparametric}, which provides an upper bound on the approximation error based on the largest eigenvalue of the covariance matrix. Portfolio optimization was subsequently performed via quadratic programming using the \texttt{solve.QP} function.

The proposed framework, referred to as \texttt{copula\_cov\_3constraint} in the performance charts, is compared against three alternatives: (i) the mean-variance model under a multivariate normality assumption using the sample covariance matrix (\texttt{data\_cov\_3constraint}), (ii) a version constrained to full investment and long-only positions (\texttt{copula\_cov\_2constraint}), and (iii) an equally weighted na\"ive portfolio (\texttt{eq\_weights}). All methods are evaluated under both daily and weekly rebalancing schemes, and performance is assessed using the average return and average Sharpe ratio over time.

\begin{figure}[h]
    \centering
    \includegraphics[width=\textwidth, height=0.45\textheight, keepaspectratio]{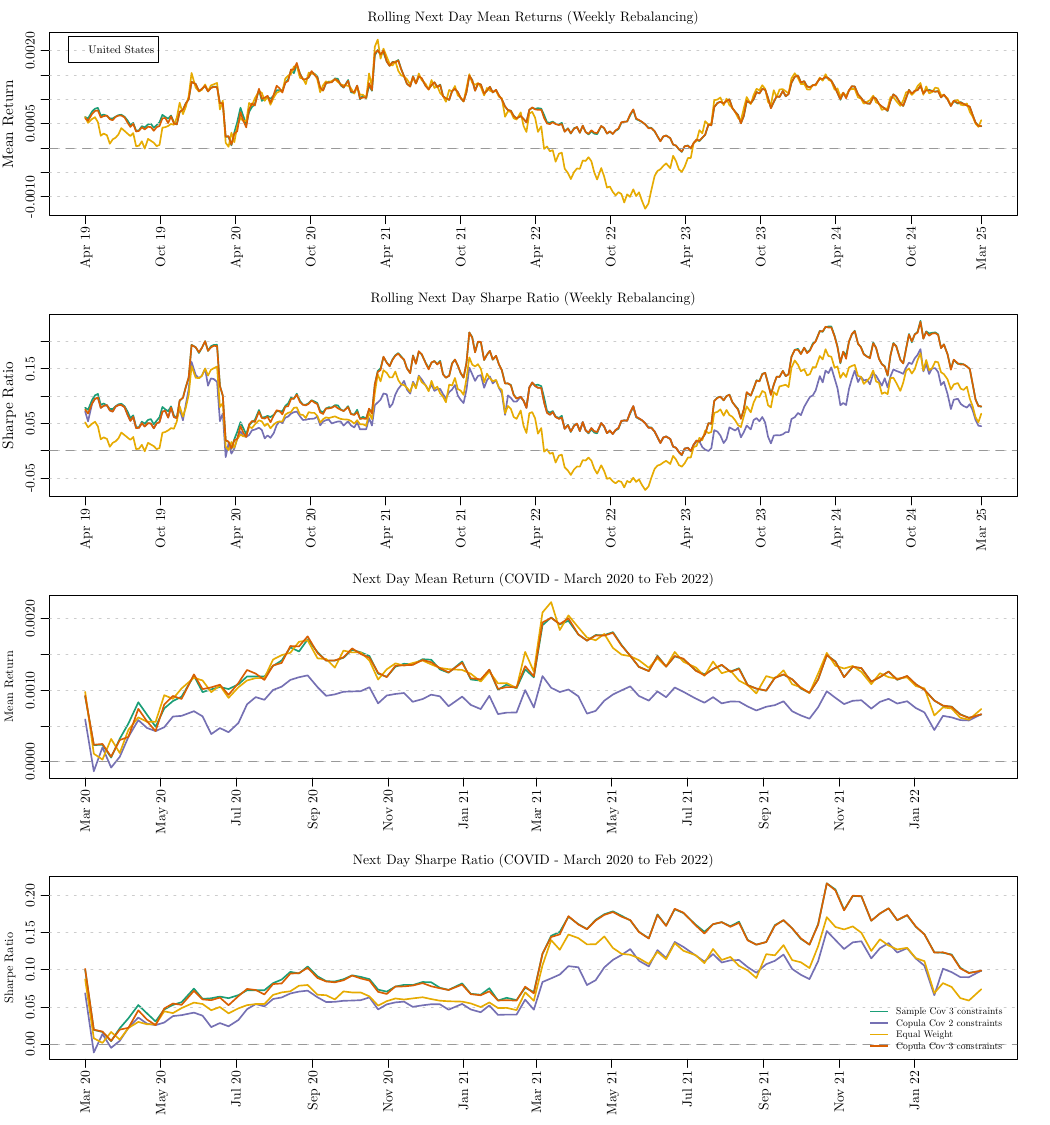} % Replace with your file name
   \caption{Rolling performance metrics for the U.S. market under weekly rebalancing (April 2019–March 2025) and during the COVID-19 period (March 2020–February 2022).
The top two panels display rolling next-day mean returns and Sharpe ratios for the U.S. market based on portfolios constructed using equal weighting, sample covariance, and copula-based covariance matrices with two and three constraints. The bottom two panels focus on the COVID-19 crisis, highlighting sharper contrasts among strategies. }
    \label{fig:sr_us}
\end{figure}

Figure~\ref{fig:sr_us} presents rolling weekly estimates of next-day mean returns and Sharpe ratios for the U.S. market from April 2019 to March 2025, with a focused view on the COVID-19 crisis period (March 2020–February 2022). Portfolios optimized using copula-based dependence structures with three constraints achieve relatively large Sharpe ratios, particularly during crisis period (e.g., mid-2020). Interestingly, in this sample, the sample covariance-based strategy performs comparably to the copula-based approach over extended periods. However, this alignment may be sample specific, as sample covariance matrices are known to be unstable under non-Gaussian return distributions or in high-dimensional settings (\citet{ledoit2004well}). By contrast, copula-based models explicitly account for tail risk and nonlinear dependence, offering greater flexibility in environments where normality assumptions do not hold.

\begin{figure}[h]
    \centering
    \includegraphics[width=\textwidth, height=0.38\textheight, keepaspectratio]{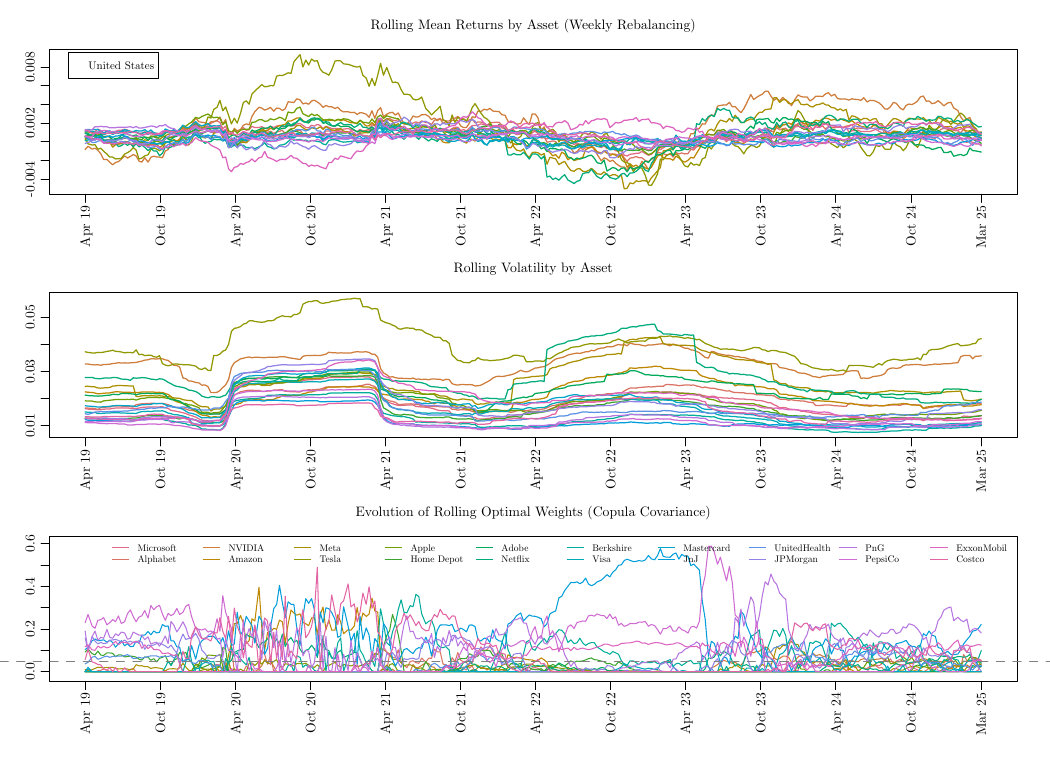} % Replace with your file name
   \caption{Rolling asset-level statistics and optimal weights for the U.S. market under weekly rebalancing (April 2019–March 2025).
The top panels present rolling mean returns and volatilities by asset, revealing heterogeneous risk-return profiles and temporal shifts, particularly during high-volatility episodes (e.g., early 2020 and 2022). The bottom panel shows the evolution of optimal portfolio weights obtained via \texttt{copula\_cov\_3constraint}. } %The observed variation in weights reflects the model’s ability to dynamically adjust allocations in response to changing asset-level risk and dependence structures.
    \label{fig:wt_us}
\end{figure}

The evolution of portfolio weights under the copula-based optimization framework (\texttt{copula\_cov\_3constraint}) exhibits distinct and interpretable shifts across three major market regimes: pre-pandemic, pandemic, and post-pandemic (Figure~\ref{fig:wt_us}). The terms pre-pandemic, pandemic, and post-pandemic refer to the periods April 2019–February 2020, March 2020–February 2022, and March 2022–March 2025, respectively. During the pre-pandemic phase, the portfolio concentrated in large-cap defensive stocks such as PepsiCo, Procter \& Gamble, and Johnson \& Johnson, consistent with a stable, low-volatility environment. In the pandemic period, the model responded dynamically to elevated uncertainty and shifting return distributions by reallocating toward resilient consumer and healthcare firms, including Amazon, Netflix, Costco, and JnJ. This period is marked by high volatility, structural breaks, and regime instability, which the copula-based framework accommodates through flexible modeling of tail dependence. In the post-pandemic phase, the portfolio progressively rotated into growth and high-beta stocks, such as Tesla, NVIDIA, and Meta, particularly during the 2023 tech-led rally. %Nonetheless, the persistence of meaningful allocations to defensives suggests a barbell strategy aimed at balancing upside capture with downside risk mitigation. 

\vspace{-.23cm}
%\vspace{-.75cm}

\begin{figure}[htbp]
    \centering
    \includegraphics[width=\textwidth, height=0.45\textheight, keepaspectratio]{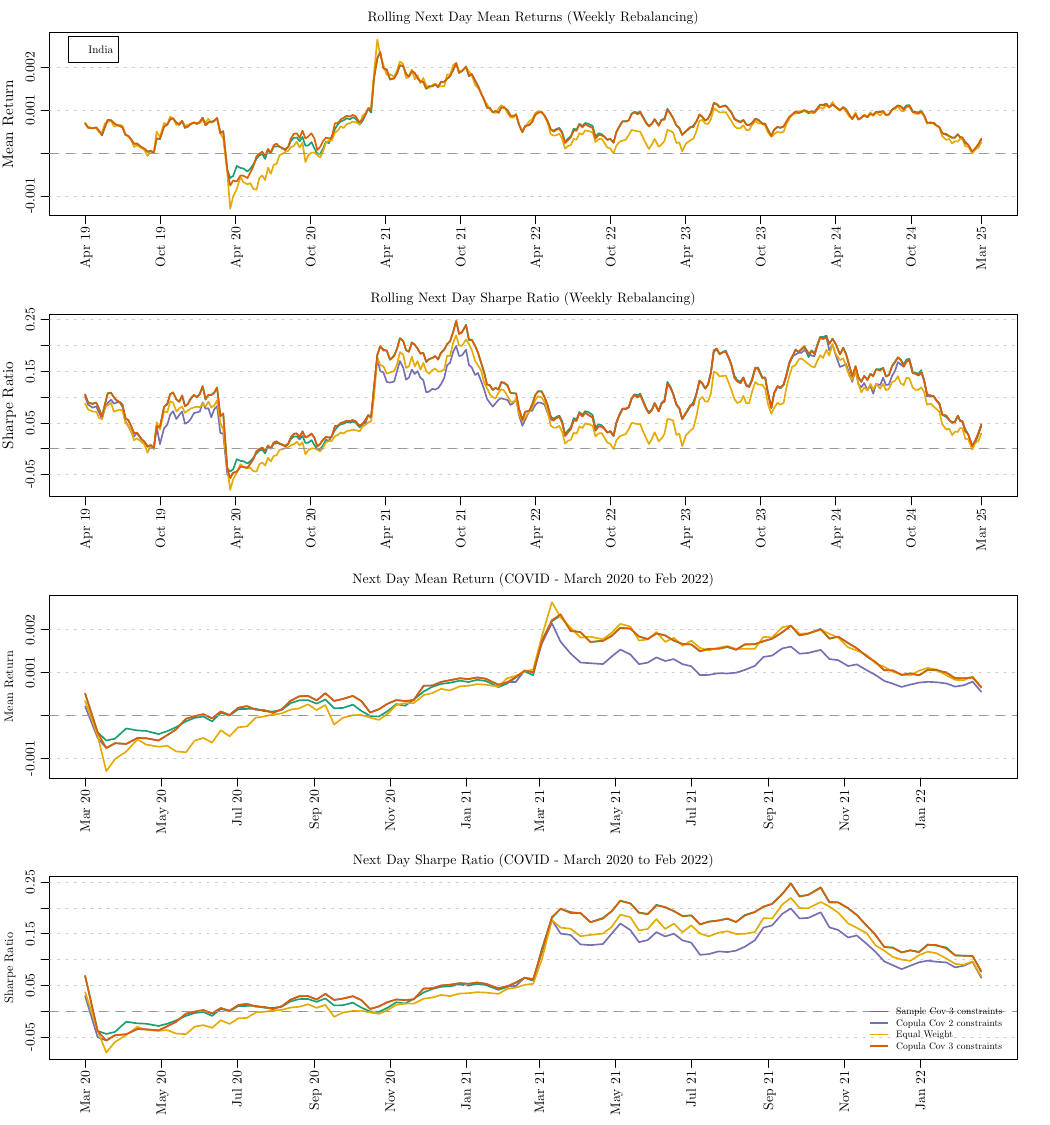} % Replace with your file name
   \caption{Rolling performance metrics for the Indian market under weekly rebalancing (April 2019–March 2025) and during the COVID-19 period (March 2020–February 2022).
The top two panels report rolling next-day mean returns and Sharpe ratios based on portfolios constructed using equal weighting, sample covariance, and copula-based covariance matrices with two and three constraints. The bottom two panels focus on the COVID-19 period, revealing that copula-based strategies outperform others in terms of risk-adjusted performance, particularly during recovery phases following market stress.}
    \label{fig:sr_in}
\end{figure}

\begin{figure}[htbp]
    \centering
    \includegraphics[width=\textwidth, height=0.38\textheight, keepaspectratio]{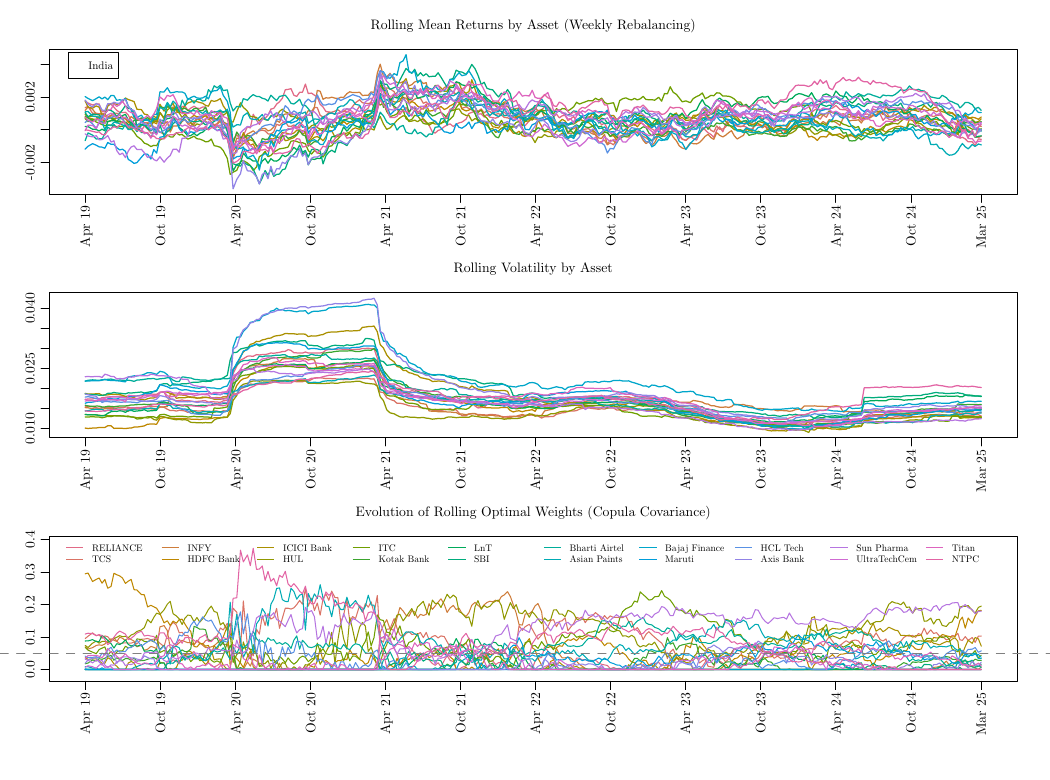} % Replace with your file name
   \caption{Rolling asset-level statistics and optimal weights for the Indian market under weekly rebalancing (April 2019–March 2025).
The top panels show rolling mean returns and volatilities by asset, capturing heterogeneous dynamics across time and securities. The bottom panel depicts the evolution of optimal portfolio weights based on \texttt{copula\_cov\_3constraint}. The time-varying allocations highlight the model’s responsiveness to shifts in risk and dependence.}
    \label{fig:wt_in}
\end{figure}

\begin{figure}[htbp]
    \centering
    \includegraphics[width=\textwidth, height=0.45\textheight, keepaspectratio]{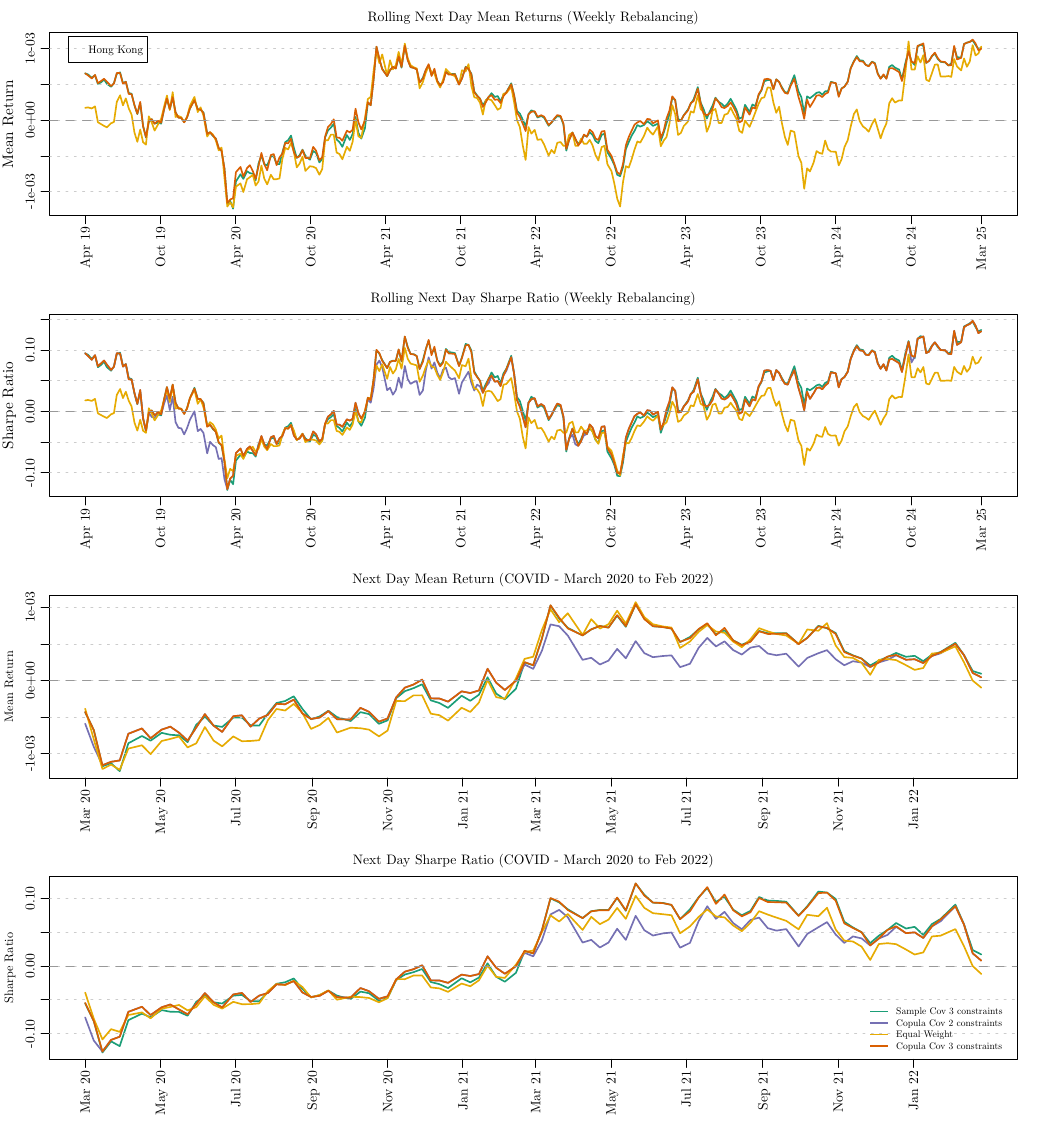} % Replace with your file name
   \caption{Rolling performance metrics for the Hong Kong market under weekly rebalancing (April 2019–March 2025) and during the COVID-19 period (March 2020–February 2022).
The top panels show rolling next-day mean returns and Sharpe ratios across portfolio strategies: equal weight, sample covariance, and copula-based covariance with two and three constraints. Copula-based portfolios with three constraints outperform consistently during volatile phases}
    \label{fig:sr_hk}
\end{figure}

\begin{figure}[htbp]
    \centering
    \includegraphics[width=\textwidth, height=0.38\textheight, keepaspectratio]{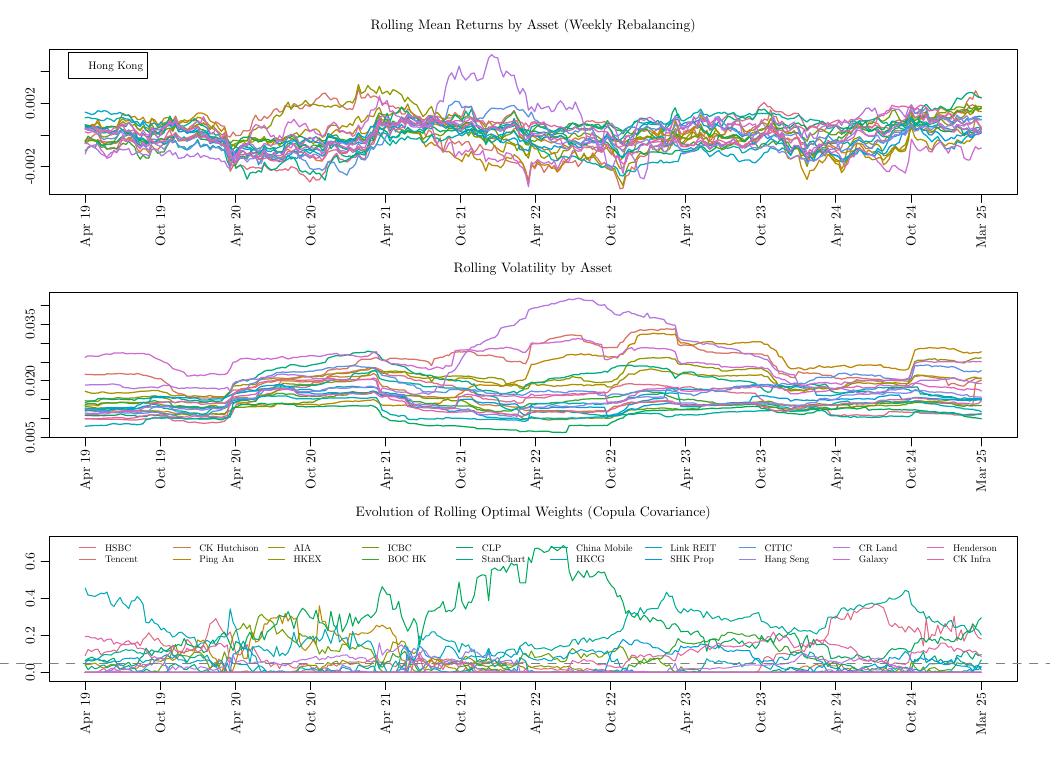} % Replace with your file name
   \caption{Rolling asset-level statistics and optimal weights for the Hong Kong market under weekly rebalancing (April 2019–March 2025).
The top panels display rolling mean returns and volatilities by asset, with sharp volatility spikes during early 2020 and late 2021. The bottom panel illustrates time-varying optimal weights based on \texttt{copula\_cov\_3constraint}.}
    \label{fig:wt_hk}
\end{figure}

\begin{figure}[h]
    \centering
    \includegraphics[width=\textwidth, height=0.45\textheight, keepaspectratio]{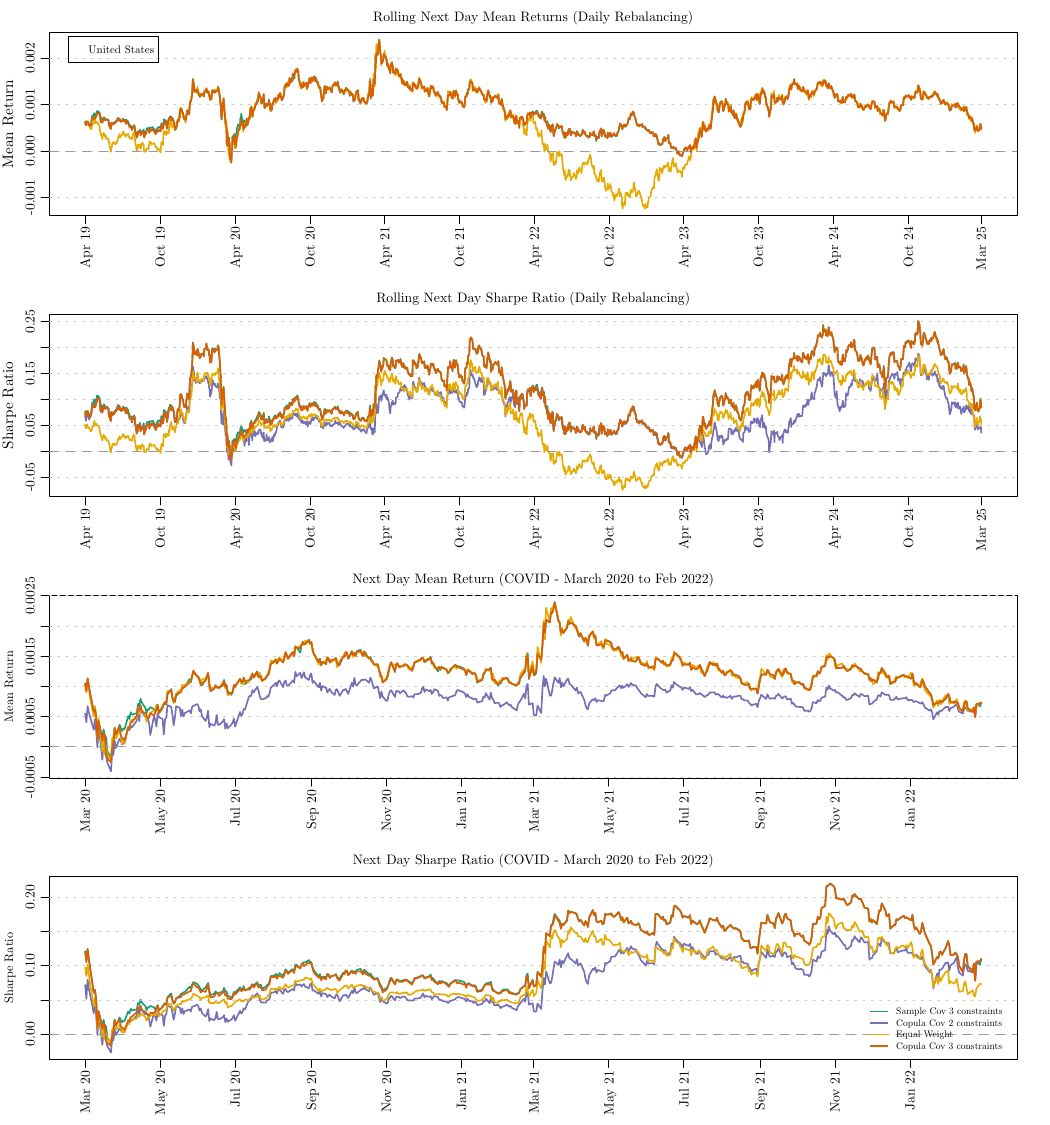} % Replace with your file name
   \caption{Rolling performance metrics for the U.S. market under daily rebalancing (April 2019–March 2025) and during the COVID-19 period (March 2020–February 2022).
The top two panels show rolling next-day mean returns and Sharpe ratios for the U.S. market using daily rebalancing across portfolio strategies: equal weight, sample covariance, and copula-based covariance with two and three constraints. Temporal patterns mirror those under weekly rebalancing, with copula-based strategies achieving more stable and elevated Sharpe ratios during high-volatility periods. The bottom two panels zoom into the COVID-19 period.}
    \label{fig:us_sr1}
\end{figure}

\begin{figure}[H]
    \centering
    \includegraphics[width=\textwidth, height=0.35\textheight, keepaspectratio]{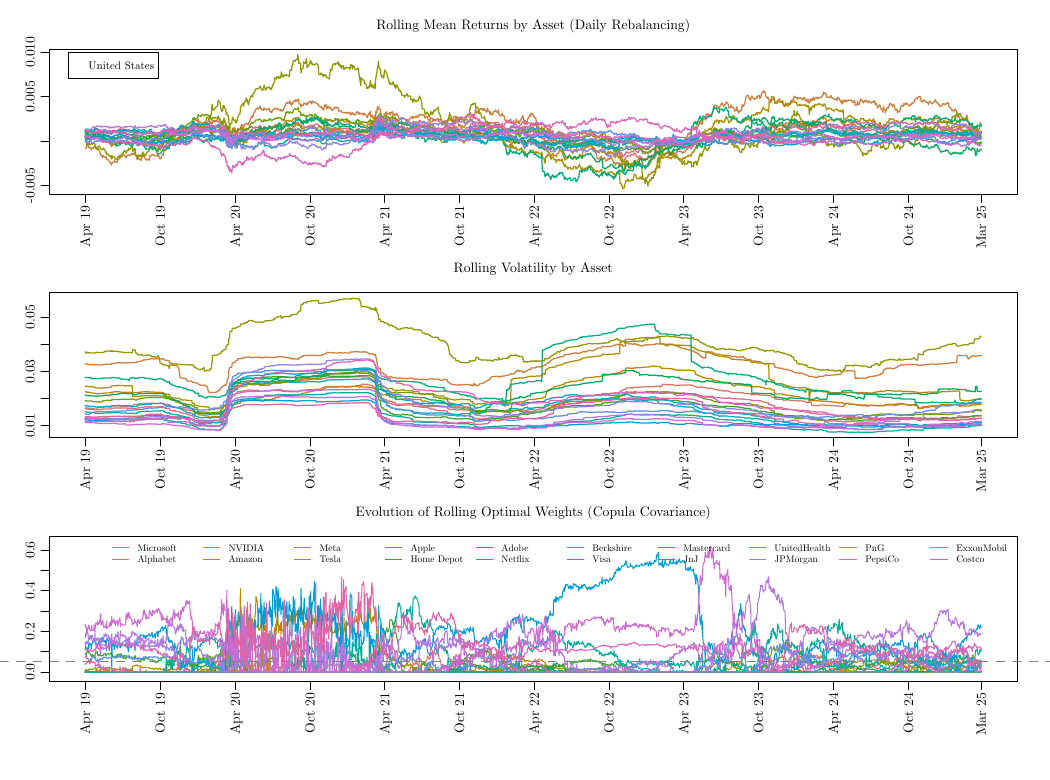} % Replace with your file name
   \caption{Rolling asset-level statistics and optimal weights for the U.S. market under daily rebalancing (April 2019–March 2025).
The top panels show rolling mean returns and volatilities by asset under daily rebalancing, capturing finer temporal fluctuations in asset behavior. The bottom panel illustrates the evolution of optimal portfolio weights derived from copula-based covariance estimation with three constraints. The dynamic reallocation patterns highlight the model’s responsiveness to short-term changes in asset-level risk and dependence. }
    \label{fig:us_wt1}
\end{figure}

\begin{figure}[h]
    \centering
    \includegraphics[width=\textwidth, height=0.45\textheight, keepaspectratio]{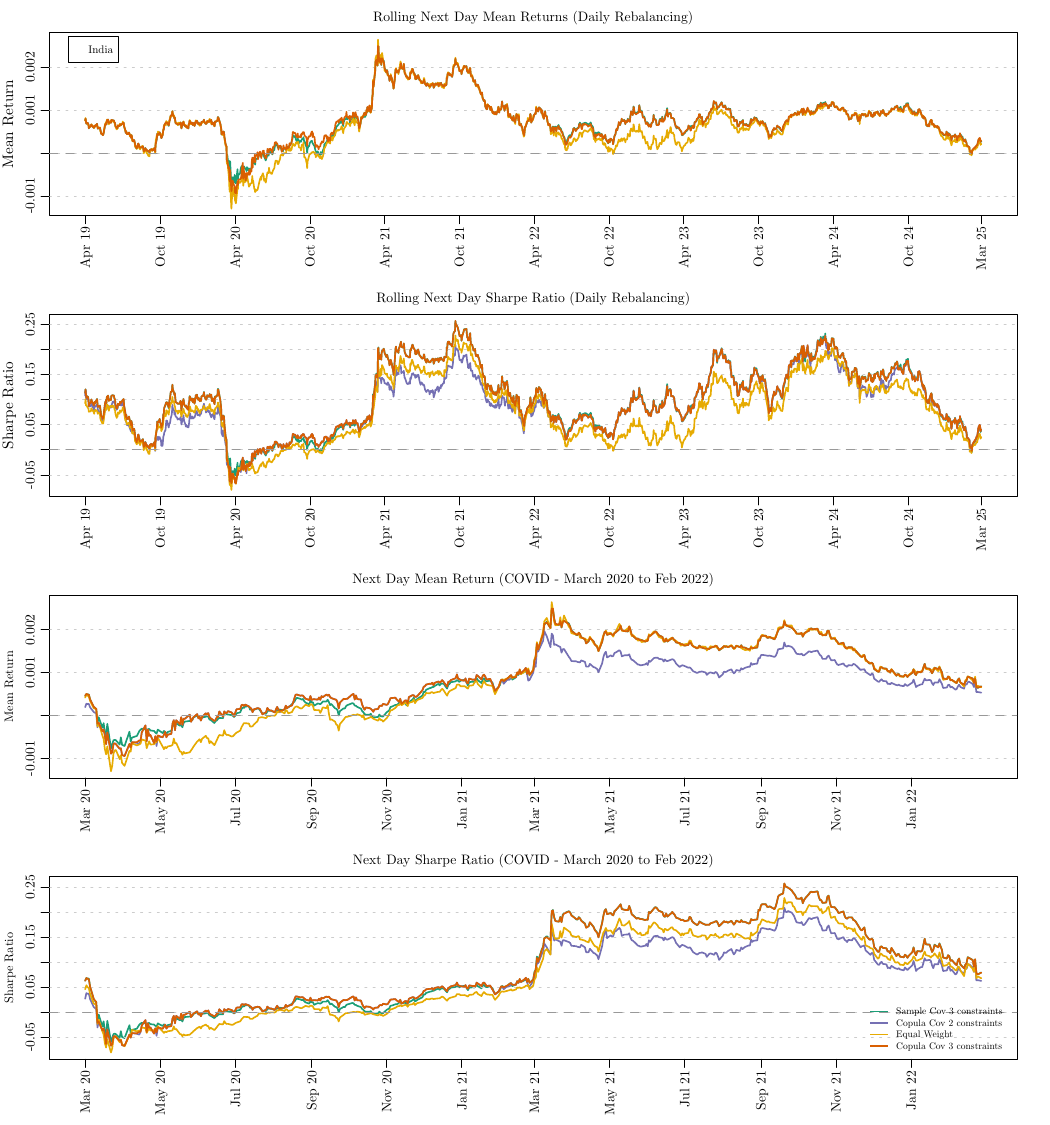} % Replace with your file name
   \caption{Rolling performance metrics for the Indian market under daily rebalancing (April 2019–March 2025) and during the COVID-19 period (March 2020–February 2022).
The top two panels report rolling next-day mean returns and Sharpe ratios based on portfolios constructed using equal weighting, sample covariance, and copula-based covariance matrices with two and three constraints. The bottom two panels focus on the COVID-19 period, revealing that copula-based strategies outperform others in terms of risk-adjusted performance, particularly during recovery phases following market stress.}
    \label{fig:in_sr1}
\end{figure}

\begin{figure}[H]
    \centering
    \includegraphics[width=\textwidth, height=0.38\textheight, keepaspectratio]{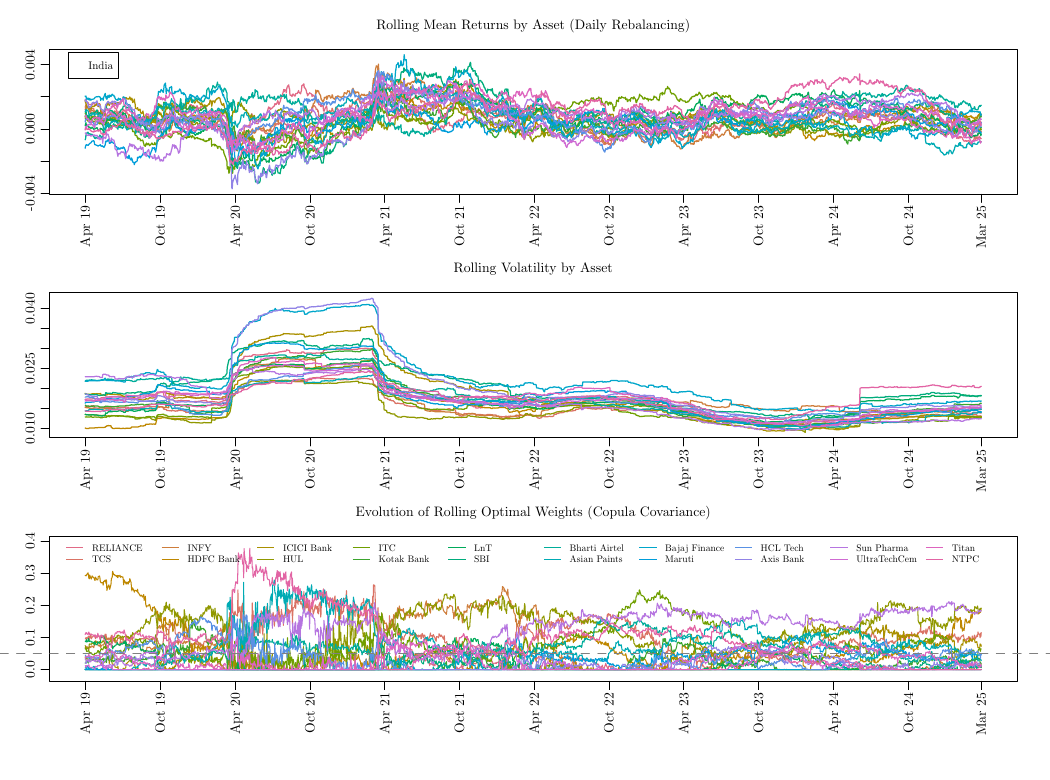} % Replace with your file name
   \caption{Rolling asset-level statistics and optimal weights for the Indian market under daily rebalancing (April 2019–March 2025).
The top panels show rolling mean returns and volatilities by asset, capturing heterogeneous dynamics across time and securities. The bottom panel depicts the evolution of optimal portfolio weights based on copula-implied covariance with three constraints.}
    \label{fig:in_wt1}
\end{figure}

\begin{figure}[h]
    \centering
    \includegraphics[width=\textwidth, height=0.45\textheight, keepaspectratio]{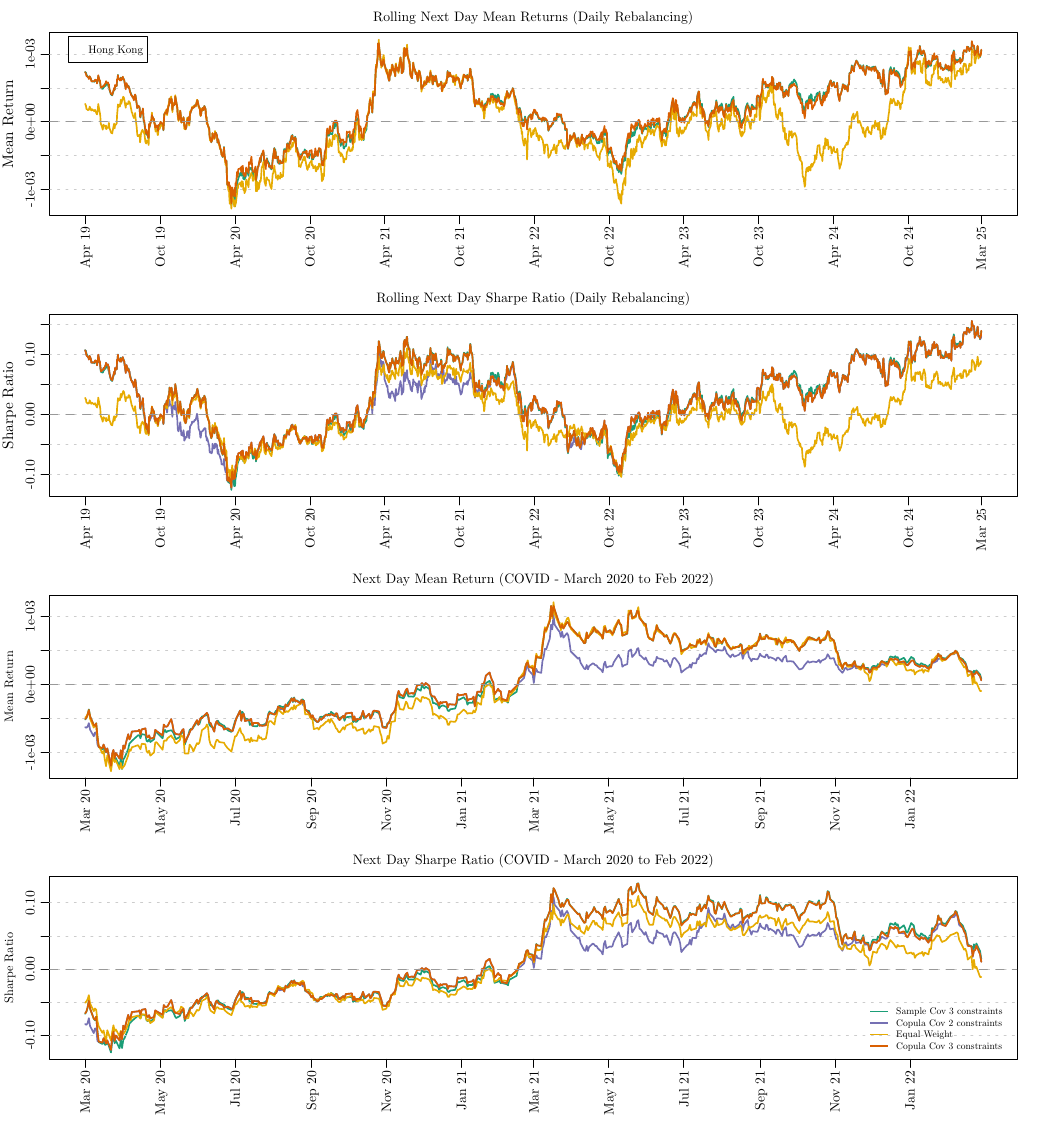} % Replace with your file name
   \caption{ Rolling performance metrics for the Hong Kong market under daily rebalancing (April 2019–March 2025) and during the COVID-19 period (March 2020–February 2022).
The top two panels display rolling next-day mean returns and Sharpe ratios across portfolio strategies: equal weight, sample covariance, and copula-based covariance estimators with two and three constraints. Copula-based portfolios with three constraints show superior stability and performance, particularly during the COVID-19 shock and recovery periods. The bottom panels zoom in on the COVID-19 phase, highlighting performance differentials under extreme market conditions.}
    \label{fig:hk_sr1}
\end{figure}

\begin{figure}[H]
    \centering
    \includegraphics[width=\textwidth, height=0.38\textheight, keepaspectratio]{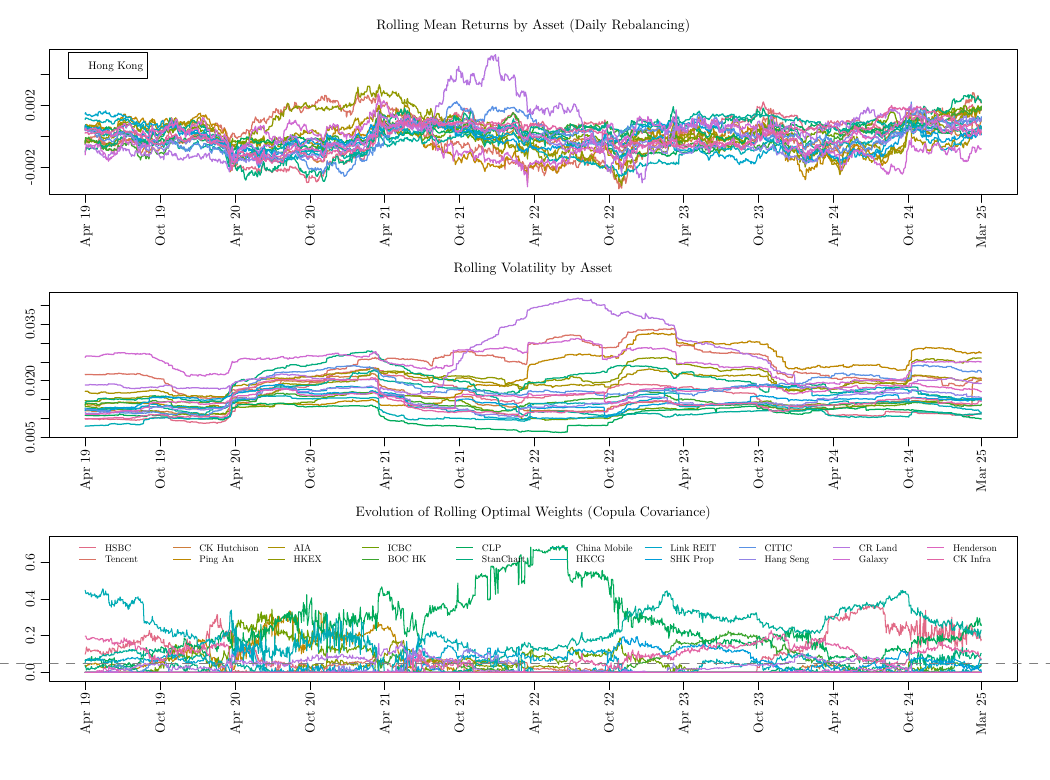} % Replace with your file name
   \caption{Rolling asset-level statistics and optimal portfolio weights for the Hong Kong market under daily rebalancing (April 2019–March 2025).
The top panels display rolling mean returns and volatilities by asset, with elevated volatility observed during periods of market stress, notably in early 2020 and late 2021. The bottom panel shows the evolution of optimal portfolio weights based on copula-implied covariance with three constraints.}
    \label{fig:hk_wt1}
\end{figure}

We replicate this analysis for the Indian and Hong Kong markets (Figures~\ref{fig:sr_in}–\ref{fig:wt_hk}). In both India and Hong Kong, the (\texttt{copula\_cov\_3constraint}) consistently outperforms equal-weighted and sample covariance benchmarks in terms of Sharpe ratio, particularly during the COVID-19 recovery period, where it delivers higher and more stable risk-adjusted returns. During the pre-pandemic phase, the Indian portfolio emphasized a mix of private financials and defensives, with top allocations to HDFC Bank, HUL, NTPC, and TCS, reflecting a stable, moderately growth-oriented stance. In the pandemic period, allocations shifted more conservatively toward utilities and non-cyclicals, led by NTPC, Asian Paints, HUL, and Sun Pharma, indicating a defensive realignment in response to heightened volatility and uncertainty. In the post-pandemic phase, the portfolio continued favoring defensive allocations: Sun Pharma, HUL, and ITC ranked highest—while partially rotating into telecom and IT names like Bharti Airtel and TCS. In the Hong Kong market, pre-pandemic phase, the portfolio was concentrated in defensives such as HKCG, HSBC, CK Infrastructure, and CLP. During the pandemic, exposure shifted further toward utilities and low-risk financials, led by CLP, HKCG, Ping An, and ICBC, reflecting heightened risk aversion. Post-pandemic, the portfolio partially rotated into telecom and diversified financials, with China Mobile, CLP, and HSBC receiving the largest weights.  These dynamics reflect a cautious recovery posture and support the presence of a barbell strategy balancing stability with selective growth exposure. These findings persist under daily rebalancing schemes (Appendix Figures~\ref{fig:us_sr1},\ref{fig:us_wt1}, \ref{fig:in_sr1},\ref{fig:in_wt1}, \ref{fig:hk_sr1} and \ref{fig:hk_wt1}), confirming the consistency of the framework in modeling time-varying risk-return trade-offs.

\begin{figure}[h]
    \centering
    \includegraphics[width=\textwidth, height=0.42\textheight, keepaspectratio]{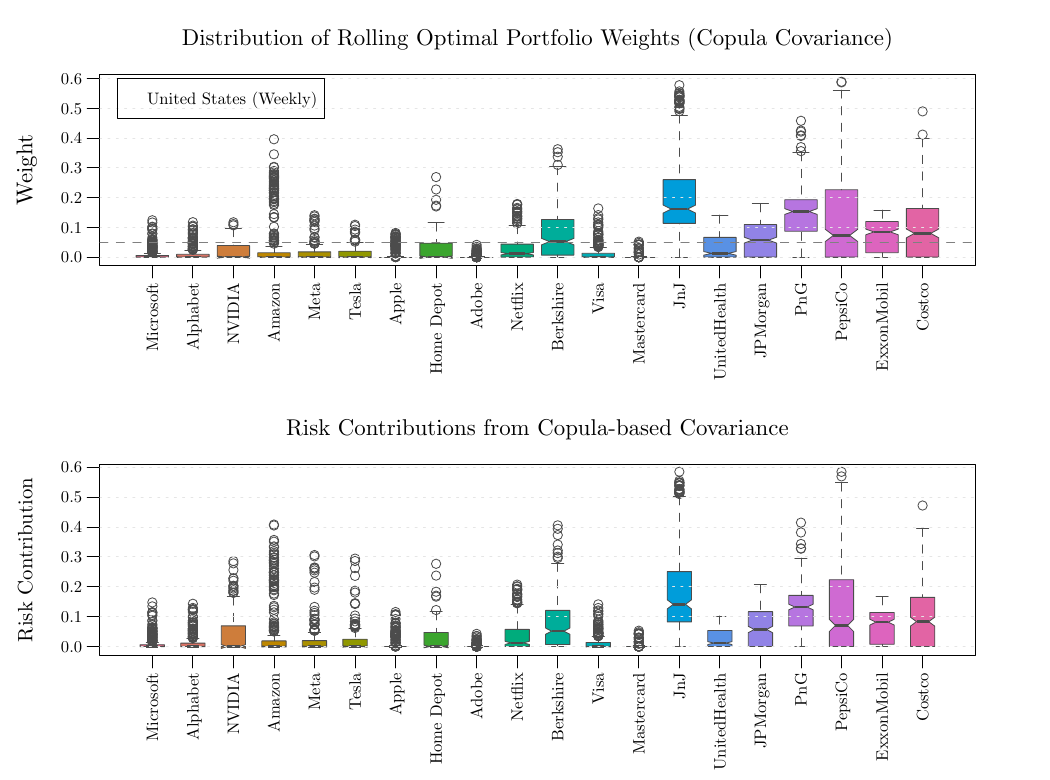} % Replace with your file name
   \caption{Time-averaged distribution of rolling optimal portfolio weights and corresponding risk contributions under the\texttt{copula\_cov\_3constraint} strategy for the U.S. market. }
    \label{fig:US_budg}
\end{figure}

\begin{figure}[!ht]
    \centering
   
        \includegraphics[width=\textwidth, height=0.42\textheight, keepaspectratio]{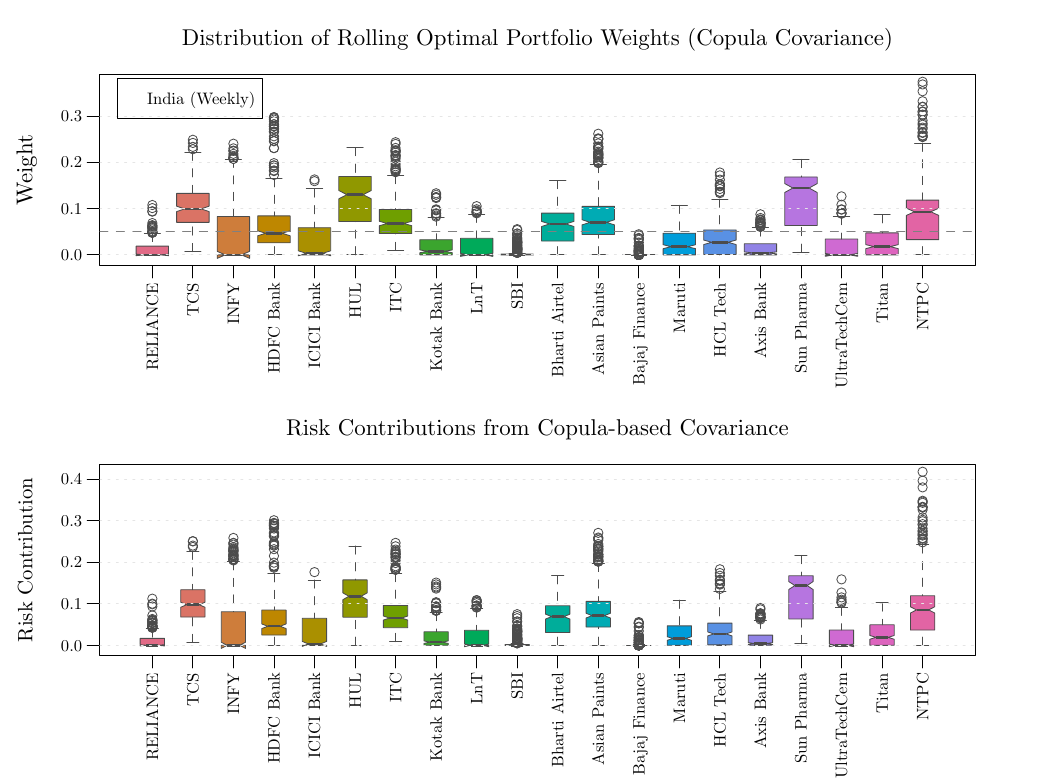}
  
    \caption{Time-averaged distribution of rolling optimal portfolio weights and corresponding risk contributions under the \texttt{copula\_cov\_3constraint} strategy for the Indian market. }
    \label{fig:IN_budg}
\end{figure}

\begin{figure}[!ht]
    \centering
   
        \includegraphics[width=\textwidth, height=0.42\textheight, keepaspectratio]{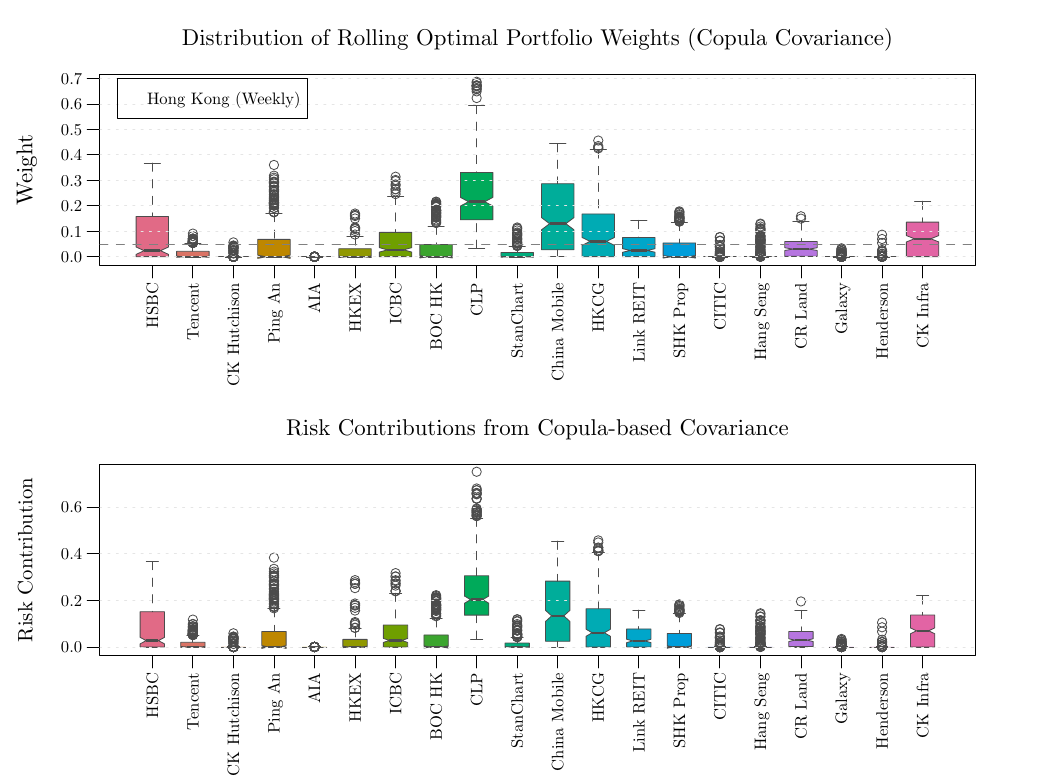}
    
    \caption{Time-averaged distribution of rolling optimal portfolio weights and corresponding risk contributions under the \texttt{copula\_cov\_3constraint} strategy for the Hong Kong market. }
    \label{fig:HK_budg}
\end{figure}

%By allowing parameters to evolve over time, dynamic dependencies linked to economic episodes are captured. We observe the model’s capacity to track evolving dependence structures and market regimes, yielding data-driven allocations under changing conditions. 
%\vspace{-4.25cm}
The time-averaged distributions of the rolling optimal weights and the corresponding risk contributions (Figures~\ref{fig:US_budg}, \ref{fig:IN_budg} and \ref{fig:HK_budg}) reveal clear market-specific patterns. The normalized risk contribution of asset \( i \) is given by:
\(
RC_i = \frac{w_i \cdot (\Sigma w)_i}{w^\top \Sigma w},
\)
where \( w_i \) is the portfolio weight of asset \( i \), \( \Sigma \) is the covariance matrix of asset returns, and \( w^\top \Sigma w \) is the total portfolio variance. This formulation ensures that \( \sum_{i=1}^N RC_i = 1 \), allowing interpretation of \( RC_i \) as the percentage contribution of asset \( i \) to total portfolio risk.
 In the U.S., weights are broadly distributed, with defensives such as PepsiCo, Procter \& Gamble, and Costco contributing consistently to total portfolio risk. In the Indian market, optimal portfolio allocations derived from copula-based covariance structures exhibit broader sectoral diversification, with noticeable weights in large-cap financials and defensives, such as TCS, HUL, ITC, Asian Paints, Sun Pharma, and NTPC. In Hong Kong, the portfolio is dominated by CLP, China Mobile, and HKCG, with these few assets accounting for a substantial share of total risk. 
These cross-market contrasts highlight how the copula framework dynamically adjusts exposure and risk budgeting in response to varying market structures and dependency patterns. 

\begin{table}[htbp]
\centering
\caption{Portfolio Performance: Copula vs. Sample Covariance (Weekly Rebalancing with Daily in Parentheses)}
\label{tab:copula-vs-sample-weekly-daily}
\resizebox{\textwidth}{!}{
\begin{tabular}{>{\raggedright\arraybackslash}p{4.6cm}ccc|ccc|ccc}
\toprule
\textbf{Metric} & 
\multicolumn{3}{c|}{\textbf{US}} & 
\multicolumn{3}{c|}{\textbf{India}} & 
\multicolumn{3}{c}{\textbf{Hong Kong}} \\
\cmidrule(lr){2-4} \cmidrule(lr){5-7} \cmidrule(lr){8-10}
& \textbf{Copula} & \textbf{Sample} & \textbf{Gain (\%)} 
& \textbf{Copula} & \textbf{Sample} & \textbf{Gain (\%)} 
& \textbf{Copula} & \textbf{Sample} & \textbf{Gain (\%)} \\
\midrule
Average Return (\%) & 
0.1204 (0.1206) & 0.1205 (0.1212) & 0.0 (-0.5) & 
0.0911 (0.0914) & 0.0894 (0.0893) & 1.9 (2.4) & 
0.0063 (0.0071) & 0.0049 (0.0059) & 28.9 (20.1) \\

Average Sharpe Ratio (\%) & 
11.1307 (11.1923) & 11.2479 (11.3327) & -1.0 (-1.2) & 
9.4066 (9.4184) & 9.3228 (9.3094) & 0.9 (1.2) & 
1.8521 (1.9273) & 1.7627 (1.8525) & 5.1 (4.0) \\

No. of Windows & 
\multicolumn{3}{c|}{101 (505)} & 
\multicolumn{3}{c|}{99 (497)} & 
\multicolumn{3}{c}{99 (494)} \\

\% Higher Return Windows & 
\multicolumn{3}{c|}{54.5 (49.3)} & 
\multicolumn{3}{c|}{69.7 (64.6)} & 
\multicolumn{3}{c}{30.3 (70.1)} \\

\% Higher Sharpe Ratio Windows & 
\multicolumn{3}{c|}{29.7 (15.8)} & 
\multicolumn{3}{c|}{59.6 (50.9)} & 
\multicolumn{3}{c}{26.3 (23.1)} \\
\bottomrule
\end{tabular}
}
%\vspace{1mm}
\begin{minipage}{\textwidth}
\footnotesize
\textit{Note:} Weekly rebalancing results are shown first, with daily rebalancing values in parentheses. `Gain' columns report the relative improvement of copula-based portfolios over sample covariance portfolios during the COVID-19 period (March 2020–February 2022).
\end{minipage}
\end{table}
During the COVID-19 crisis period, we evaluate the relative performance of the (\texttt{copula\_cov\_3constraint}) strategy against the sample covariance-based approach. Table~\ref{tab:copula-vs-sample-weekly-daily} presents a comparative analysis of portfolio performance under copula-based and sample covariance-based optimization frameworks across the US, India, and Hong Kong markets. The copula-based method demonstrates superior performance in most scenarios, particularly under weekly rebalancing. Notably, the Hong Kong market exhibits the largest relative gains, with improvements of 28.9\% in mean return and 5.1\% in Sharpe ratio, highlighting the method’s advantage in capturing asymmetric and tail-dependent relationships. Across all three markets, the copula approach achieved higher average Sharpe ratios and returns in a substantial proportion of rolling windows, for instance, outperforming in 54.5\% (US), 69.7\% (India), and 30.3\% (Hong Kong) of windows in terms of return, and in 29.7\%, 59.6\%, and 26.3\% of windows in terms of Sharpe ratio, respectively. 
In the Hong Kong market, although the number of outperforming windows is lower, the relative gains in those periods are substantially larger, indicating that when the copula-based strategy does outperform, it does so by a significant margin. These results shows that copula-based allocations, especially during volatile periods such as the COVID-19 crisis, by effectively modeling non-linear dependencies and extreme co-movements often missed by sample covariance-based models.

\subsection{Skewed Generalized \texorpdfstring{$t$}{t} Distribution for Asset Returns}
We employed the Skewed Generalized \(t\) (SGT) distribution to model individual asset returns, which were then input into a copula function to characterize their dependence structure. The SGT distribution is well-suited to financial returns because it can flexibly capture skewness, heavy tails, and excess kurtosis, thus providing a suitable alternative to conventional symmetric distributions. It features distinct parameters for location (mean), scale (standard deviation), and shape (skewness and tail thickness). %Moreover, several well-known heavy-tailed or skewed distributions, such as the skewed generalized error distribution (SGED), skewed \(t\) (ST), skewed Laplace, generalized error distribution (GED), skewed normal (SN), Student’s \(t\), skewed Cauchy (SCauchy), and even the normal distribution arise as special or limiting cases of the SGT. Primary references for this distribution in finance include \cite{theodossiou1998sgt}, \cite{Theodossiou2016Skewness}.

Using the \texttt{sgt} package in \texttt{R} (\cite{sgt}), the parameters of the SGT distribution can be estimated by maximum likelihood using the function \texttt{sgt.mle(x, mu, sigma, lambda, p, q, mean.cent = TRUE, var.adj = TRUE)}. For our analysis, we initialized the parameter estimates with the sample mean (\(\mu\)) and sample standard deviation (\(\sigma\)) of the data, along with small values for the shape parameters \((p=q=2)\) to capture the heavy tails that are typically observed in financial returns. The \texttt{mean.cent = TRUE} option ensures that \(\mu\) corresponds to the actual mean of the SGT distribution, while \texttt{var.adj = TRUE} adjusts \(\sigma\) to represent the standard deviation. Details on the functional form and derivations are in the \texttt{sgt} package vignette, which provides a detailed implementation guide. After estimating the parameters, we first plot the fitted density for rolling windows to visualize the evolution of the marginal distributions over time, capturing the dynamic nature of financial returns. 
 
 Figure \ref{fig:roll_den1} presents the SGT-fitted overlaid rolling density plots for the Netflix and Costco US market stocks. The variations in these plots show the dynamic nature of financial returns over time, potentially reflecting shifts in market conditions, investor sentiment, or external factors. Although densities exhibit changes over time, these changes are not overly drastic. The SGT family appears to effectively account for different distributional shapes characterized by high kurtosis and skewness. Similar behavior is observed in the Indian and Hong Kong markets. %(Figures \ref{fig:insgt} and \ref{fig:hksgt}). editt..
 
To assess the goodness of fit of the SGT distribution, we use the p-value from the AD test, which evaluates how closely the estimated SGT distribution represents the observed stock returns. Since the true distribution of the data is unknown, we consider the fitted SGT distribution as a proxy for the true distribution. A high p-value suggests that the observed stock returns align well with the theoretical SGT distribution, validating its suitability for modeling financial return data. The P values were calculated in R using the \texttt{ADGofTest} package (\cite{adgof}), which applies the AD test and an approximate p-value method from \cite{mars}. All p-values exceed 0.2, providing strong evidence that the SGT distribution effectively models the marginal densities of financial returns. Summary measures, including time-averaged \( p \)-values and estimated asymmetry parameters from the SGT fit, are presented as box plots in Appendix Figures~\ref{fig:sgt_us}, \ref{fig:sgt_in}, and \ref{fig:sgt_hk}.


\section{Discussion}
The proposed semiparametric model which combines nonparametrically estimated copula with parametrically estimated marginals allows all parameters to dynamically evolve over time making it suitable for predictive inference. Applied to rolling windows of financial returns from the USA, India and Hong Kong economies, this approach addresses the limitations of traditional models that rely on parametric assumptions.   %By accounting for asymmetry, heavy tails, and cross-correlated asset prices, the proposed method offers a robust solution for optimizing diverse portfolios in an interconnected financial markets.
%In this work, we proposed a flexible and adaptive framework to compute the next-day portfolio net worth using Markowitz's mean-variance model. 
%*extend results bit more..)
The skewed generalized t distribution captures skewness and kurtosis often observed in financial returns, while the empirical beta copula, as a nonparametric estimator, accommodates arbitrary and multivariate dependence structures. The copula-implied dependence structure enables adaptive, data-driven portfolio decisions across heterogeneous market environments. The model effectively tracks evolving inter-asset dependencies and regime shifts, translating structural changes into rebalanced allocations over time. 

By allowing parameters to evolve over time, the proposed framework (\texttt{copula\_cov\_3constraint}) captures dynamic dependencies linked to economic episodes, offering substantial improvements in modeling time-varying, nonlinear, and asymmetric relationships relative to other approaches. It demonstrates consistent outperformance during periods of financial stress, where classical models often fail to reflect joint tail risk and dynamic correlation shifts. Each rolling window, comprising marginal SGT estimation, empirical beta copula construction, and simulation-based optimization with \( m = 10^5 \) samples, completes in approximately 20--25 seconds on a standard Intel Core i7 (2.90\,GHz, 16\,GB RAM) machine. The method entails a modest computational burden but provides greater flexibility and reliability in crisis regimes. Its modular design is readily parallelizable, supporting scalable implementation in high-frequency or large-universe settings. 
From the market analyses, the model performed reasonably well across all three markets by identifying dynamically evolving weights that optimize portfolio returns. The results indicate that optimally weighted portfolios from \texttt{copula\_cov\_3constraint} consistently outperform equal-weighted portfolios, with higher weights allocated to high-return assets, reflecting their importance in the portfolio. Future work will involve exploring other risk measures (e.g., VaR and CVaR) using the proposed semiparametric dynamic models. The R code supporting the results of this study will be made publicly available on the first author’s GitHub repository upon acceptance of the paper.  %(\url{https://shorturl.at/5amxi})

%\textcolor{red}{GHOSH COMMENT: Have you created a GitHub page? It is not necessary for submission but eventually we will need one after the paper is accepted.}
%Future work can extend it to other asset classes, investigate alternative optimization criteria, and explore different rebalancing strategies to further enhance its applicability. 
%\bibliographystyle{plainnat}
%\bibliography{references}  
% Comment this out
% \bibliographystyle{apalike} 
% \bibliography{ref}

% Add this instead:

\newpage
\section*{Appendix}
%\vspace{-.5cm}
\begin{table}[htbp]
\centering
\scriptsize
\caption{Summary Statistics of Selected US Stocks (April 2018 -- March 2025)}
\label{tab:us-summary-stats}
%\resizebox{\textwidth}{!}{%
\begin{tabular}{lllccl}
\toprule
\textbf{Ticker} & \textbf{Company Name} & \textbf{Sector} & \textbf{Mean Return (\%)} & \textbf{Mean SD (\%)} & \textbf{Skewness} \\
\midrule
MSFT & Microsoft Corporation & Information Technology & 0.09 & 1.81 & -0.29 \\
GOOGL & Alphabet Inc. (Class A) & Communication Services & 0.06 & 1.95 & -0.24 \\
NVDA & NVIDIA Corporation & Information Technology & 0.17 & 3.29 & -0.24 \\
AMZN & Amazon.com, Inc. & Consumer Discretionary & 0.06 & 2.16 & -0.14 \\
META & Meta Platforms, Inc. & Communication Services & 0.07 & 2.67 & -1.34 \\
TSLA & Tesla, Inc. & Consumer Discretionary & 0.16 & 4.06 & -0.07 \\
AAPL & Apple Inc. & Information Technology & 0.10 & 1.93 & -0.23 \\
HD & The Home Depot, Inc. & Consumer Discretionary & 0.05 & 1.70 & -1.43 \\
ADBE & Adobe Inc. & Information Technology & 0.03 & 2.33 & -0.83 \\
NFLX & Netflix, Inc. & Communication Services & 0.07 & 2.83 & -2.18 \\
BRK-B & Berkshire Hathaway Inc. & Financials & 0.06 & 1.30 & -0.21 \\
V & Visa Inc. & Financials & 0.06 & 1.64 & -0.07 \\
MA & Mastercard Incorporated & Financials & 0.07 & 1.84 & 0.03 \\
JNJ & Johnson \& Johnson & Health Care & 0.03 & 1.22 & -0.33 \\
UNH & UnitedHealth Group & Health Care & 0.06 & 1.80 & -0.53 \\
JPM & JPMorgan Chase \& Co. & Financials & 0.06 & 1.86 & -0.02 \\
PG & Procter \& Gamble Company & Consumer Staples & 0.05 & 1.27 & -0.03 \\
PEP & PepsiCo, Inc. & Consumer Staples & 0.03 & 1.31 & -0.57 \\
XOM & Exxon Mobil Corporation & Energy & 0.05 & 1.95 & -0.16 \\
COST & Costco Wholesale Corporation & Consumer Staples & 0.10 & 1.45 & -0.53 \\
\bottomrule
\end{tabular}%
%}
\end{table}
%\vspace{-.5cm}

\FloatBarrier
\begin{table}[htbp]
\centering
\scriptsize
\caption{Summary Statistics of Selected Indian Stocks (April 2018 -- March 2025)}
\label{tab:india-summary-stats}
%\resizebox{\textwidth}{!}{%
\begin{tabular}{lllccl}
\toprule
\textbf{Ticker} & \textbf{Company Name} & \textbf{Sector} & \textbf{Mean Return (\%)} & \textbf{Mean SD (\%)} & \textbf{Skewness} \\
\midrule
RELIANCE.NS   & Reliance Industries Limited         & Oil, Gas \& Consumable Fuels        & 0.07 & 1.82 & 0.05  \\
TCS.NS        & Tata Consultancy Services Limited   & Information Technology              & 0.06 & 1.53 & -0.02 \\
INFY.NS       & Infosys Limited                     & Information Technology              & 0.07 & 1.74 & -0.64 \\
HDFCBANK.NS   & HDFC Bank Limited                   & Financial Services                  & 0.04 & 1.58 & -0.35 \\
ICICIBANK.NS  & ICICI Bank Limited                  & Financial Services                  & 0.10 & 1.95 & -0.49 \\
HINDUNILVR.NS & Hindustan Unilever Limited          & Fast Moving Consumer Goods          & 0.04 & 1.44 & 0.73  \\
ITC.NS        & ITC Limited                         & Fast Moving Consumer Goods          & 0.04 & 1.56 & -0.65 \\
KOTAKBANK.NS  & Kotak Mahindra Bank Limited         & Financial Services                  & 0.04 & 1.77 & -0.28 \\
LT.NS         & Larsen \& Toubro Limited            & Construction                        & 0.06 & 1.74 & -0.62 \\
SBIN.NS       & State Bank of India                 & Financial Services                  & 0.07 & 2.07 & -0.39 \\
BHARTIARTL.NS & Bharti Airtel Limited               & Telecommunication                   & 0.09 & 1.89 & 0.31  \\
ASIANPAINT.NS & Asian Paints Limited                & Consumer Durables                   & 0.04 & 1.60 & -0.41 \\
BAJFINANCE.NS & Bajaj Finance Limited               & Financial Services                  & 0.09 & 2.35 & -0.95 \\
MARUTI.NS     & Maruti Suzuki India Limited         & Automobile and Auto Components      & 0.02 & 1.86 & -0.21 \\
HCLTECH.NS    & HCL Technologies Limited            & Information Technology              & 0.08 & 1.72 & -0.10 \\
AXISBANK.NS   & Axis Bank Limited                   & Financial Services                  & 0.05 & 2.19 & -1.65 \\
SUNPHARMA.NS  & Sun Pharmaceutical Industries Ltd   & Healthcare                          & 0.07 & 1.76 & -0.02 \\
ULTRACEMCO.NS & UltraTech Cement Limited            & Construction Materials              & 0.06 & 1.75 & -0.16 \\
TITAN.NS      & Titan Company Limited               & Consumer Durables                   & 0.07 & 1.82 & -0.31 \\
NTPC.NS       & NTPC Limited                        & Power                               & 0.07 & 1.77 & -0.43 \\
\bottomrule
\end{tabular}%
%}
\end{table}
\FloatBarrier
%\vspace{-56cm}
%\vspace{-.5cm}
\begin{table}[H]
\centering
\scriptsize
\caption{Summary Statistics of Selected Hong Kong Stocks (April 2018 -- March 2025)}
\label{tab:hk-summary-stats}
%\resizebox{\textwidth}{!}{%
\begin{tabular}{lllccl}
\toprule
\textbf{Ticker} & \textbf{Company Name} & \textbf{Sector} & \textbf{Mean Return (\%)} & \textbf{Mean SD (\%)} & \textbf{Skewness} \\
\midrule
0005.HK  & HSBC Holdings plc                                   & Financial Services        & 0.03   & 1.50 & -0.22 \\
0700.HK  & Tencent Holdings Limited                            & Communication Services    & 0.02   & 2.40 &  0.41 \\
0001.HK  & CK Hutchison Holdings Limited                       & Diversified Holdings              & -0.02  & 1.54 &  1.32 \\
2318.HK  & Ping An Insurance Co. of China                      & Financial Services        & -0.01  & 2.20 &  0.33 \\
1299.HK  & AIA Group Limited                                   & Financial Services        & -0.0002& 1.96 &  0.001 \\
0388.HK  & HK Exchanges and Clearing Ltd.                      & Financial Services        & 0.03   & 2.05 &  0.47 \\
1398.HK  & Industrial \& Commercial Bank of China              & Financial Services        & 0.02   & 1.35 &  0.38 \\
2388.HK  & BOC Hong Kong (Holdings) Ltd.                       & Financial Services        & 0.01   & 1.43 &  0.19 \\
0002.HK  & CLP Holdings Limited                                & Utilities                 & 0.005  & 1.10 & -0.23 \\
2888.HK  & Standard Chartered PLC                              & Financial Services        & 0.03   & 1.93 & -0.18 \\
0941.HK  & China Mobile Limited                                & Communication Services    & 0.03   & 1.39 &  0.70 \\
0003.HK  & HK \& China Gas Co. Ltd.                            & Utilities                 & -0.02  & 1.26 & -1.41 \\
0823.HK  & Link REIT                                           & Real Estate Investment    & -0.01  & 1.50 & -0.37 \\
0016.HK  & Sun Hung Kai Properties Ltd.                        & Real Estate               & -0.01  & 1.46 & -0.07 \\
0267.HK  & CITIC Limited                                       & Diversified Holdings              & 0.02   & 1.85 & -0.03 \\
0011.HK  & Hang Seng Bank Limited                              & Financial Services        & -0.01  & 1.45 & -0.10 \\
0836.HK  & China Resources Power Holdings                      & Utilities                 & 0.04   & 2.46 &  0.14 \\
0027.HK  & Galaxy Entertainment Group Ltd.                     & Consumer Discretionary    & -0.04  & 2.53 & -0.19 \\
0012.HK  & Henderson Land Development Co. Ltd.                 & Real Estate               & -0.01  & 1.60 & -0.03 \\
1038.HK  & CK Infrastructure Holdings Ltd.                     & Infrastructure            & 0.004  & 1.39 & -0.56 \\
\bottomrule
\end{tabular}%
%}
\end{table}

%\subsection*{United States}

\begin{figure}[H]
    \centering
        \includegraphics[width=\textwidth, height=0.42\textheight, keepaspectratio]{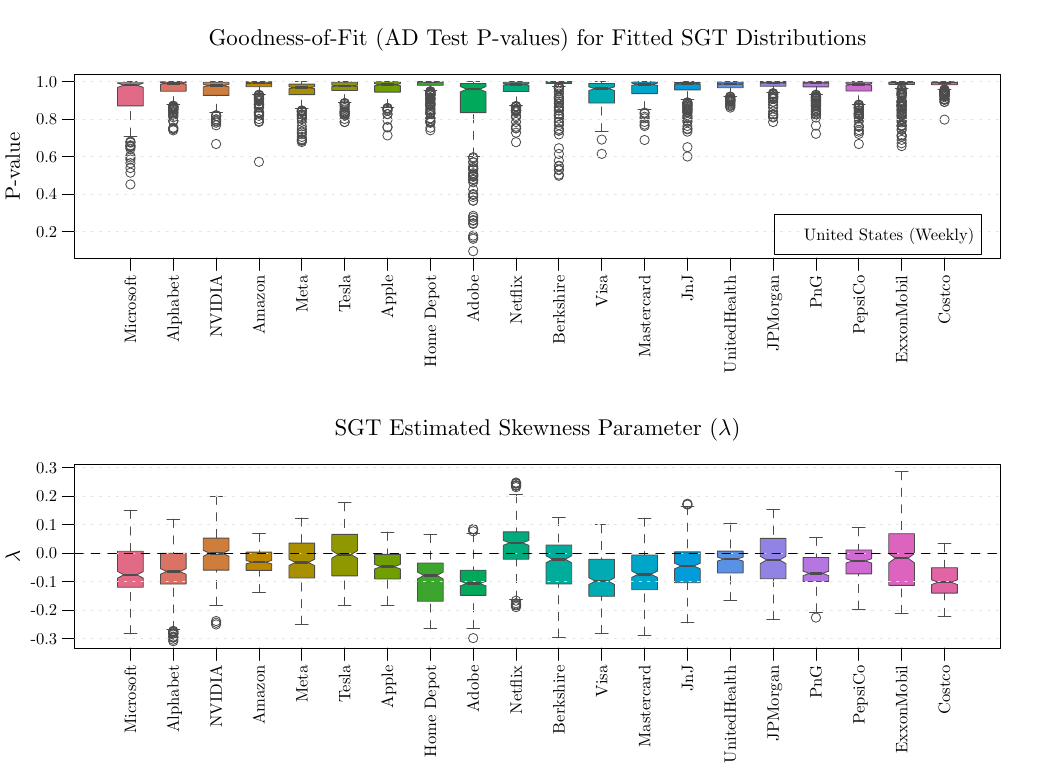}
    \caption{Goodness-of-fit and skewness estimates for SGT models — United States.
P-values from the AD test and estimated skewness parameters $(\lambda)$ for weekly return distributions fitted using the SGT-distribution across U.S. equities. The results demonstrate strong model fit and highlight moderate to negative skewness in several technology and consumer stocks.}
    \label{fig:sgt_us}
\end{figure}
%\subsection*{India}


\begin{figure}[!ht]
    \centering
   
        \includegraphics[width=\textwidth, height=0.42\textheight, keepaspectratio]{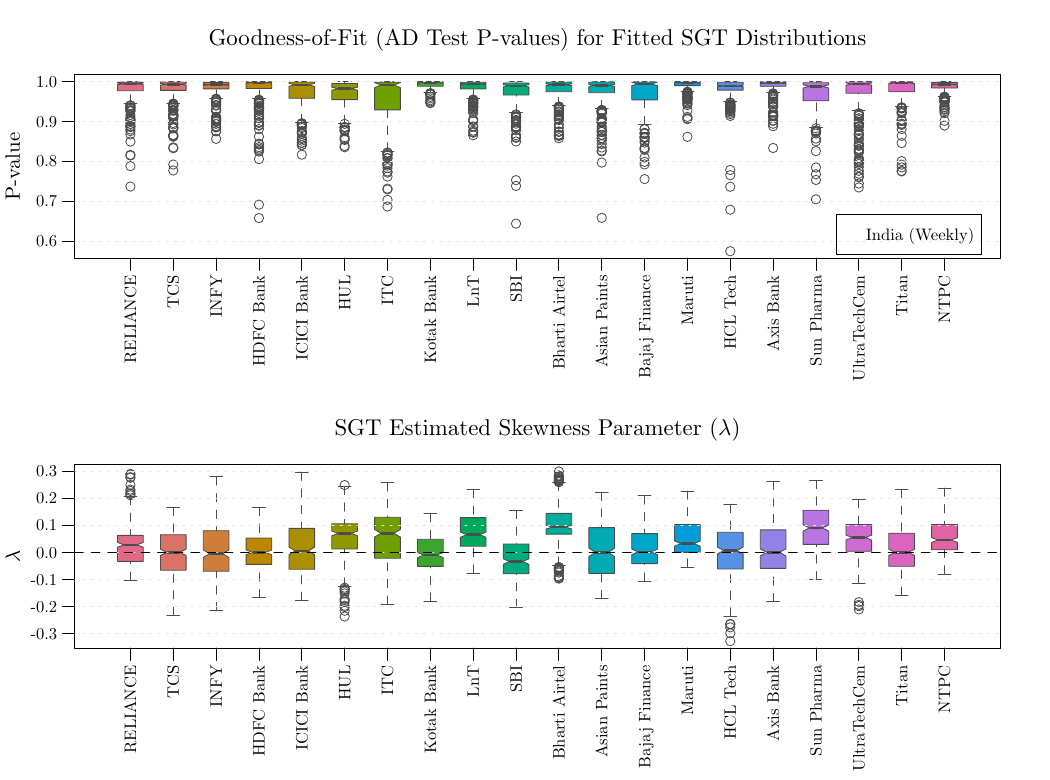}

    \caption{Goodness-of-fit and skewness estimates for SGT models — India.
AD test p-values and skewness estimates for SGT-fitted weekly return distributions across major Indian stocks. Most assets exhibit good distributional fit and mildly positive skewness, suggesting upside asymmetry among defensives and large-cap financials.}
    \label{fig:sgt_in}
\end{figure}

%\subsection*{Hong Kong}

\begin{figure}[!ht]
    \centering
   
        \includegraphics[width=\textwidth, height=0.42\textheight, keepaspectratio]{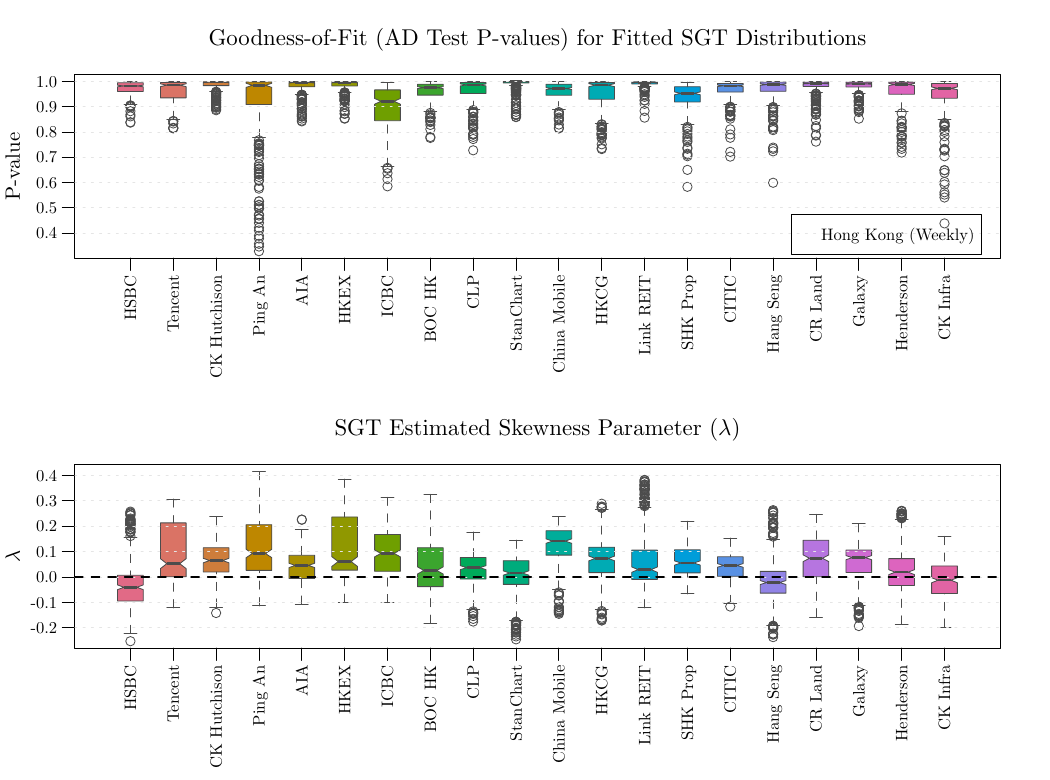}
  
    \caption{Goodness-of-fit and skewness estimates for SGT models — Hong Kong.
SGT model evaluation for Hong Kong equities using AD test p-values and estimated $\lambda$ parameters. The fitted models show consistent goodness-of-fit, with moderate positive skewness evident in utilities and telecom sectors, indicating asymmetric tail behavior typical of the region’s market structure.}
    \label{fig:sgt_hk}
\end{figure}

\end{document}